\documentclass[11pt,a4paper]{article} 
\pdfoutput=1
\usepackage[utf8x]{inputenc}     
\usepackage{jheppub}
\usepackage{enumerate}
\usepackage{epsfig} 
\usepackage{float} 
\usepackage[caption = false]{subfig}
\usepackage{wasysym} 
\usepackage{mathrsfs} 
\usepackage{amsfonts} 
\usepackage{amsbsy} 
\usepackage{amscd} 
\usepackage{cancel}
\usepackage{multirow} 
\usepackage{tikz}
\usepackage{color}
\usepackage{slashed}
\usepackage{array}
\usepackage{tablefootnote}

\usetikzlibrary{arrows,positioning,shapes.geometric} 
\usepackage[compat=1.1.0]{tikz-feynman}          
\usepackage[font=small,labelfont=bf]{caption}
\usepackage{diagbox}
\usepackage{tabularx}
\usepackage{soul}  
\usepackage{listings} 
\usepackage{orcidlink}           
\DeclareUnicodeCharacter{2212}{\textendash}    
\usepackage{amsthm}

\title{Precision prediction of a democratic up-family philic KSVZ axion model at the LHC}  
\author[a,b]{Anupam Ghosh\,\orcidlink{0000-0003-4163-4491},}
\author[a]{and Partha Konar\,\orcidlink{0000-0001-8796-1688}}

\affiliation[a]{Theoretical Physics Division, Physical Research Laboratory, Shree Pannalal Patel Marg, Ahmedabad, 380009, Gujarat, India}
\affiliation[b]{Discipline of Physics, Indian Institute of Technology, Palaj, Gandhinagar, 382424, Gujarat, India}

\emailAdd{anupam@prl.res.in}
\emailAdd{konar@prl.res.in}

\abstract{
In this work, we study the $SU(2)_L$ singlet complex scalar extended KSVZ model that, in addition to providing a natural solution to the strong-CP problem, furnishes two components of dark matter that satisfy observer relic density without fine-tuning the model's parameters. A colored vector-like quark (VLQ) is naturally present in the KSVZ axion model, providing a rich dark matter and collider phenomenology. In this extended model, scalar dark matter interacts with the Standard Model up-type quarks (up, charm, top) through VLQ. We explore the possibility of democratic Yukawa interaction of the VLQ with all up-type quarks and scalar dark matter candidate. We also employ next-to-leading order NLO-QCD correction on dominant production channels for VLQ pair production to study a unique search at the LHC, generating a  pair of boosted tops with sizeable missing transverse momentum. Such corrections are significant and reduce factorization and renormalization scale uncertainties substantially. The NLO fixed order results are matched with the Pythia8 parton shower. After being pair-produced, each VLQ decays into a dark matter and a top quark. We conducted a multivariate analysis using jet substructure variables of boosted top fatjets with a significant missing transverse momentum signal. This analysis allows us to explore a substantial parameter space of this model at the 14 TeV LHC.
}

\preprint{\today}


\begin{document}
\maketitle

\newpage

\section{Introduction}
\label{Intro_5}

The Standard Model (SM) of particle physics harbours several shortcomings despite providing the most successful theory of the underlying nature of fundamental particles and their interactions, verified in many low to high-energy experiments consistently delivering excellent agreement. Among many such issues relating to some internal inconsistencies to the inability to explain some of the observations, significant flaws like the strong-CP problem \cite{Cheng:1987gp, Baluni:1978rf,Dine:2000cj}, the existence of dark matter (DM) \cite{Sofue:2000jx, Clowe:2006eq}, baryon-antibaryon asymmetry of the universe \cite{Riotto:1999yt, Dine:2003ax}, neutrino masses \cite{Super-Kamiokande:1998kpq, SNO:2002tuh, K2K:2002icj}, etc. are the ones which generated lots of attention in motivating to go beyond the Standard Model (BSM) and their searches at different experiments. 

Due to the nontrivial QCD vacuum structure, the QCD Lagrangian includes an $\theta$-dependent term that does not obey parity and time reversal invariance. The physically observable parameter is the effective $\overline{\theta}$ term when considered in the basis where the quark mass matrix is real and diagonal. The $SU(3)_C$ symmetry of the SM allows a term like $\theta\dfrac{g_S^2}{32\pi^2} \tilde{G}_{\mu \nu}^a G^{a,\mu \nu}$, where $G^{a, \mu \nu}$ is the gluon field strength tensor, and $g_S$ is the strong coupling constant. $\tilde{G}_{\mu \nu}^a=\epsilon_{\mu\nu\rho\sigma} G^{a,\rho\sigma}$, $\epsilon_{\mu\nu\rho\sigma}$ is the antisymmetric tensor with $\epsilon^{0123}=+1$. This term contributes to the neutron electric dipole moment, and the experimental measurement \cite{PhysRevLett.124.081803} constraints the parameter $\overline{\theta}\leq 10^{-10}$. These two parameters $\overline{\theta}$ and $\theta$ are related through quark field chiral rotation. Since $\overline{\theta}\rightarrow 0$ does not promote any symmetry of the theory, one would anticipate $\overline{\theta}\sim\mathcal{O}(1)$, which is known as a strong charge-parity (CP) problem~\cite{Cheng:1987gp, Baluni:1978rf, Dine:2000cj}. 
Roberto Peccei and Helen Quinn proposed a classic resolution to this critical issue in 1977  by extending the SM with a global Peccei-Quinn (PQ) symmetry, expected to be broken spontaneously at a scale far larger than the Electroweak (EW) scale. The breaking of $U(1)_{PQ}$ predicts the existence of a pseudo-Goldstone particle, also known as the QCD axion. It is even more interesting to note that although QCD axion is not entirely stable, it can have a lifetime comparable to the age of the Universe, thanks to the sizeable breaking scale, to play the role of dark matter \cite{Peccei:1977hh,Peccei:1977ur}. Hence such models can concurrently explain the presence of the DM in the Universe while also solving the strong-CP problem. 

After breaking $U(1)_{PQ}$ symmetry, one may formulate an axion-gluon effective Lagrangian
where $F_a$ is the Peccei-Quinn breaking scale and this Lagrangian appears to be 
\begin{equation}
\mathcal{L} = \mathcal{L}_{\text{QCD}}+ \theta \dfrac{g_S^2}{32\pi^2} \tilde{G}_{\mu \nu} G^{\mu \nu}+\dfrac{1}{2}\partial_\mu a \partial^\mu a - \dfrac{g_S^2}{32\pi^2 F_a} a(x) \tilde{G}_{\mu \nu} G^{\mu \nu}.
\label{Eq.axion_5}
\end{equation} 
$\mathcal{L}_{\text{QCD}}$ represents the QCD Lagrangian, and $a(x)$ is the axion field. After minimizing the axion potential, one can write $\overline{\theta}=\theta - \dfrac{1}{ F_a}<a(x)>$. As a result, the vacuum expectation value of the axion field cancels the original $\theta$, and the strong CP issue is resolved. 
%

Kim-Shifman-Vainshtein-Zakharov's (KSVZ) model \cite{Kim:1979if, Shifman:1979if} provides an exciting phenomenological implementation among different existing models. The KSVZ is a UV complete axion model that solves the strong CP problem and provides the QCD axion as a dark matter candidate.
This model includes a PQ-breaking complex scalar ($\eta$) that is singlet under the SM gauge groups, as well as a vector-like quark (VLQ) $\Psi = \Psi_L+\Psi_R$ that is $SU(3)_C$ color triplet but $SU(2)_L$ singlet. The interaction term between VLQ and the scalar is $f_\Psi \eta \overline{\Psi}_L \Psi_R + h.c$, where $f_\Psi$ is the interaction strength. The scalar field can be written as,
\begin{equation}
\eta = \frac{1}{\sqrt{2}} \Bigl(F_a+\sigma_0(x)\Bigr)~e^{\frac{ia(x)}{F_a}}.
\label{Eq.eta_5}
\end{equation}
Here $F_a$ is the PQ breaking scale, $\sigma_0(x)$ and $a(x)$ are the radial mode and the axion field, respectively. Following the symmetry breaking, VLQ gains mass proportionate to the PQ breaking scale, and they are thus isolated from the axion field and may be safely integrated out. This leads to an axion-gluon effective Lagrangian (last term of Equation~\ref{Eq.axion_5}). Axion dynamically takes the value corresponding to the minimum axion potential and cancels the original $\theta$, and thus the strong-CP problem is resolved. The QCD axion mass calculated at next-to-leading order (NLO) \cite{GrillidiCortona:2015jxo} is
\begin{equation}
m_a = 5.70 \Bigl( \frac{10^{12}~\text{GeV}}{F_a} \Bigr)~\mu eV~.
\label{Eq:axion_mass}
\end{equation}
Axion can decay into gluons because of the above effective axion-gluon interaction (Eq. \ref{Eq.axion_5}), and its decay rate is inversely proportional to the PQ breaking scale. Therefore, if the braking scale is tuned correctly, the axion lifetime can be larger than the age of the Universe and behaves as a DM.

The KSVZ model is a renormalized QCD-axion model that inherently contains a color fermion (VLQ) and a Pecci-Quinn symmetry-breaking complex scalar, $\eta$, whose phase part is the QCD-axion. It solves two outstanding problems of the Standard Model of particle physics: the strong-CP problem and Dark Matter. The axion can behave as dark matter and provide the correct relic density of DM, as measured by the Planck collaboration \cite{Akrami:2018odb}, for a specific $F_a$ breaking scale. $F_a$ can pick any value between $10^{10}~GeV ~\leq F_a \leq 10^{12}~GeV$, with the lower limit arising from the supernova cooling data \cite{PhysRevLett.60.1793} and the upper limit arising from the axion overproduction. We extend this setup by introducing a complex scalar, $S$, a singlet under the SM gauge groups, and we do not need a fine-tuned value of $F_a$ to achieve the observed dark matter relic density. This gives us a two-component dark matter scenario, the axion and the scalar DM collectively account for the observed relic. It is also worth noting that even after the breaking of the PQ symmetry, the KSVZ model leaves a residual $\mathbb{Z}_2$ symmetry intact, stabilizing the lightest component of the scalar $S$. Now, we aim to investigate the dark matter and collider phenomenology within this singlet scalar extended KSVZ framework.

VLQ interacts with Standard Model quarks and the scalar $S$ in the present configuration. The hypercharge of VLQ is determined by the kind (up or down) of the SM quarks considered. Given that we are considering up-type quarks, the hypercharge of VLQ is $\frac{2}{3}$. 
Since DM interacts with the SM quarks through VLQ, additional direct detection channels open up, such as VLQ-mediated s and t-channel elastic scattering diagrams between the SM-quark and scalar DM. Moreover, the VLQ and its interaction with the SM quarks also affect the LHC phenomenology. VLQ decays into an SM-quark and a missing DM particle after being produced at the LHC. As a result, multijet plus missing transverse momentum may be employed as a possible probe. If the mass difference between VLQ and DM is more than the top quark mass, VLQ can be probed from its decay into top quarks, along with a sizeable missing transverse energy from dark matter in the final state.  

The Yukawa interaction takes the form $f_{i}S \overline{\Psi}_L {u_i}_R +~h.c$, where $u_R$ denotes right-handed up-type SM-quarks with $i = u,~ c,~ t$. We first consider the parameter spaces that yield the correct relic density and are permissible from other experimental observations such as direct detection (DD), collider data, etc., with equal (democratic) coupling strengths $f_u=f_c=f_t$ at all three generations. Interestingly, one would find that the flavor constraint strongly disfavors this democratic option, although such models can be allowed from observed relic density and all other constraints (please follow the flavor constraints part in Section \ref{model_5}). {\sl The flavor constraint requires either or both \cite{Ghosh:2022rta} lighter flavor couplings ($f_u$, $f_c$) tiny to be allowed}. Instead of taking a tiny value for $f_c$, we set $f_c=0$ while keeping the other two democratic $f_u=f_t=f$ \footnote{The reason why we do not favor $f_u=0$ is addressed in the flavor constraints section.}. We found that the parameter spaces that support correct relic density hold considerably higher $f$ values. The choice of this parameter generates a very different physics outlook both in DM phenomenology and collider constraints compared to the scenario with large $f_t$ ($\sim 1$) and negligible couplings for the other two generations ($f_c, f_u$) \cite{Ghosh:2022rta}.
We summarise some of the salient features of the present work as follows.
\begin{itemize}
\item The QCD axion solves the strong CP problem. It also provides a suitable dark matter candidate for a choice of specific breaking scale.
\item In this multicomponent scenario, the scalar candidate and axion share the desired fraction to satisfy the observed dark matter relic for breaking scale anywhere between $10^{10}$ to $10^{12}$ GeV.
\item More interestingly, there is no requirement to introduce an ad-hoc discrete symmetry, such as $\mathbb{Z}_2$, to ensure the stability of the scalar dark matter, as this $\mathbb{Z}_2$ symmetry remains intact after the breaking of the PQ symmetry.
\item Because its coupling is inversely proportional to the PQ breaking scale $F_a$, the QCD axion remains inaccessible at the LHC. Instead, we explore this model by investigating the coloured vector-like quark at the LHC.
\item After production, VLQs decay into SM quarks associated with the scalar DM, making the boosted top-fat jets with significant missing energy an excellent channel to probe at the LHC.
\item We also computed the next-to-leading order (NLO) correction for VLQ pair production, which is significant.
\item The current analysis utilizes jet-substructure variables and employs the multivariate analysis technique to enhance the reach of the probe.
\end{itemize}

The present study investigates the reach of this compelling parameter space at the 14 TeV LHC. We especially employ next-to-leading order (NLO) correction for VLQ pair production for precise computation and match the NLO fixed-order result to the Pythia8 parton shower. The partonic leading-order (LO) cross section has the order $\sigma(p p \rightarrow \Psi \bar{\Psi})_\text{LO}=\mathcal{O}(\alpha_S^2)+\mathcal{O}(f^4)+\mathcal{O}(f^2\alpha_S)$. The dominant contribution arises from the pure QCD sector ($\mathcal{O}(\alpha_S^2)$), and we consider $\alpha_S$ corrections of those processes. The integrated NLO-K factor for pure QCD processes is around 1.3, which means $30\%$ enhancement over the LO cross section. We also observe that the differential distributions change significantly, and the theoretical scale uncertainties reduce considerably.

Another interesting point is that the scalar DM parameter spaces that provide the correct relic density while simultaneously being allowed by direct detection constraints differ dramatically. In the ref. \cite{Ghosh:2022rta}, for example, when the mass of the scalar DM is greater than the mass of the top quark ($m_t$), DM annihilates into $t\bar{t}$ through VLQ exchange t-channel, giving the correct relic density when $f_t\sim 1$ while the other two couplings $f_u, f_c$ are tiny. A tiny $f_u$ is required since it has to be allowed from the direct detection experimental constraint. In the current study, when the DM is heavier than the top quark, DM annihilation into $t\bar{u},~\bar{t}u$ contributes the most in relic density, followed by the annihilation into $t\bar{t}$ final state. Likewise, allowed parameter space can neither support arbitrarily large coupling $f$ ($=f_u=f_t$) from direct detection nor the too-small value of it to obtain correct relic density. Therefore, their interplay remains vital for selecting the available parameter spaces. In contrast to ref. \cite{Ghosh:2022rta}, only a tiny model space remains when the mass difference between the scalar DM and the VLQ ($\Delta M_{\Psi S_1}$) is smaller than $m_t$, as direct detection constraints prohibit such scenarios with a high $f$ value, despite achieving the correct relic density.

After the pair production of VLQs at the LHC, each VLQ decays into the top quark associated with the scalar since the majority of parameter spaces are present when $\Delta M_{\Psi S_1}>m_t$. The branching ratio $\text{BR}(\Psi\rightarrow t S)<0.5$ and depends on the coupling $f$. Our signal comprises two boosted top-like fatjets and missing transverse energy (MET). We consider all the SM background processes that mimic the signal. We do a sophisticated multivariate analysis of two top fatjets plus a significant MET signal using the Boosted Decision Tree (BDT) algorithm. The available higher-order QCD cross section is used to normalize all the background processes. The parameter spaces of this model are shown to be well within the scope of the 14 TeV LHC with 300 $\text{fb}^{-1}$ luminosity.

The paper is structured as follows. Section \ref{model_5} introduces our model and a brief outlook on different theoretical and experimental constraints. The dark matter phenomenology of this model is discussed in Section \ref{DMphenomenology_5}. Section \ref{nlops_5} demonstrates the impact of NLO+PS calculations, the differential k-factor, and the scale uncertainty of NLO+PS compared to LO+PS. Section \ref{collider_5} displays our collider analysis technique. Finally, we summarise our findings in Section \ref{sec:conclsn_5}. 


\section{The extended KSVZ model and Constraints}
\label{model_5}

As expressed in the introduction, the KSVZ model contains a complex scalar $\eta$ (Equation \ref{Eq.eta_5}), which breaks the PQ-symmetry spontaneously, and color triplet vector-like quark ($\Psi$). The generic Lagrangian of the KSVZ model can be expressed as,
\begin{equation}
\mathcal{L}^{\text{KSVZ}}=~ \partial_\mu \eta^\dagger \partial^\mu \eta + \bar{\Psi} i \gamma^\mu D_\mu \Psi -(f_\Psi \eta \bar{\Psi}_L\Psi_R+h.c)  - \lambda_\eta (|\eta|^2-F_a^2/2)^2~.
\label{EQ. KSVZ_5}
\end{equation}
We extend the model by a complex singlet scalar field,\footnote{An $SU(2)_L$ doublet extension of the KSVZ model is also viable within the WIMP-axion framework. For a comprehensive study of the dark matter phenomenology and an in-depth investigation of this doublet extended model at the LHC, see reference \cite{Ghosh:2024boo}.} $S$ with the same PQ charge as $\Psi_L$ (see Eq. \ref{Lag.VLQ_5}). The scalar can be written as,
\begin{equation}
S=\frac{S_1+iS_2}{\sqrt{2}}~.
\label{Eq.scalar_5}
\end{equation}
%
A residual $\mathbb{Z}_2$ symmetry will remain intact after the spontaneously broken PQ-symmetry of the KSVZ model that stabilizes the lightest component of the singlet scalar. Hence this takes a role of a second dark matter candidate in this theory as a weakly interacting massive particle (WIMP). 
We consider VLQ to have non-zero hypercharge so that the scalar DM can interact with the SM up-type quarks through the mediator VLQ. Such a construction with the up-type quark instead of the down-type has an interesting consequence. Pair of scalar DM can annihilate into a top-pair or a single top associated with a light quark ($q=u,c$) through t-channel VLQ mediated diagrams (see Fig.~\ref{annihilation_5}) depending on whether DM mass satisfies such kinematic limit and provide the observed relic density.
Therefore, aside from the Higgs portal, when the dark matter has a mass nearly half that of the Higgs boson, it annihilates into the Standard Model particles through an on-shell Higgs resonance, providing the correct relic abundance. A heavier DM, on the other hand, opens up new parameter spaces and also yields the observed relic density. Another compelling reason comes from the interesting phenomenological viewpoint. Heavy VLQ can be produced copiously at a high-energy collider, which in turn decays into the top quark and a missing DM particle whenever kinematically feasible. This can result in a unique topology in the LHC search, such as the possibility of a boosted top-fatjets along with a sizable amount of missing transverse momentum from dark matter. This is a likely scenario if one notes down the present constraint on VLQ mass, which is already in the vicinity of the TeV scale, and such a heavy state would naturally produce decay products which are sufficiently boosted. 

The interaction terms of VLQ are given below.
\begin{equation}
\mathcal{L}^{\text{VLQ}}=~ -(f_{i}S \overline{\Psi}_L {u_i}_R + f_\Psi \eta \overline{\Psi}_L \Psi_R + h.c.), \qquad \{ i=u,~c,~t\}
\label{Lag.VLQ_5}
\end{equation}
The full scalar potential of the model can be written as,
%
\begin{equation}
\begin{split}
V=&~ \lambda_H (|H|^2 - v_H^2/2)^2 +\lambda_\eta (|\eta|^2-F_a^2/2)^2+\lambda_{\eta H} (|H|^2 - v_H^2/2) (|\eta|^2-F_a^2/2)  \\
&+\mu_S^2|S|^2+\lambda_S |S|^4 +\lambda_{SH} |H|^2|S|^2 + \lambda_{S \eta } |\eta|^2 |S|^2 +[\epsilon_S \eta^* S^2 +h.c].
\end{split}
\label{potential_5}
\end{equation}
Where $v_H$ is the VEV of the SM Higgs potential, the second term is the Mexican-hat potential of the KSVZ model.
%
Because of the third term, the mixing between the SM Higgs field and the radial part of the PQ-braking scalar $\eta$ happens, which leads to a non-diagonal mass matrix. After the diagonalization of the mass matrix, one can get the physical masses, which can be expressed below.
\begin{equation}
M_{h,\sigma}^2 = (\lambda_H v^2 +\lambda_\eta F_a^2 ) \pm \sqrt{(\lambda_H v^2 -\lambda_\eta F_a^2 )^2+ F_a^2 v^2 \lambda_{\eta H}^2} 
\label{phys_mass_5}
\end{equation}
The physical scalar Higgs boson mass is set to $M_h=125$ GeV, and $v_H$ is equated with the SM electroweak VEV, $v_H=v=246$ GeV. However, $\lambda_H$ differs from the SM value when $\lambda_{\eta H}\neq 0$. It is evident from the above equation that the mass of the radial excitation of the field $\eta$ is $M_\sigma \sim F_a$ for $\lambda_\eta \sim 1$. Since $F_a\sim 10^{11}~\text{GeV}$, $M_\sigma$ is immensely high, resulting in its decoupling from the remaining particle spectrum within the model. $\lambda_S$ represents the self-interaction strength between scalars and plays no role in LHC and dark matter phenomenology.

The masses of the different components of the scalar and the VLQ can be expressed as,
\begin{equation}
M_{S_{1,2}}^2=~\frac{1}{2}(2\mu_S^2+v_H^2\lambda_{SH}+ F_a^2 \lambda_{S \eta } \mp 2\sqrt{2}\epsilon_s F_a), \qquad M_{\Psi}=~f_{\Psi}\frac{F_a}{\sqrt{2}}.
\label{DM_mass,VLQ_mass}
\end{equation}
We consider $F_a$ and $M_{\Psi}$ as the independent parameters, and therefore $f_{\Psi}$ is the dependent parameter. Hence one requires a minuscule  coupling, $f_{\Psi}\leq 10^{-6}$ in eq.~\ref{DM_mass,VLQ_mass}, given that $F_a\geq 10^{10}$ GeV, and the mass of VLQ is in TeV range.  Appendix \ref{appen_60} demonstrates a simple realization of TeV scale VLQ without necessitating such a minuscule coupling of $f_{\Psi}$. Without losing generality, we can assume that $S_1$ is lighter than $S_2$ (alternatively, $\epsilon_s$ takes a real positive value). This lighter component $S_1$ is the scalar DM. 

KSVZ model with the extended complex scalar is widely studied in the literature~\cite{Dasgupta:2013cwa,Chatterjee:2018mac}; it was noted that a certain amount of fine-tuning is unavoidable to achieve $M_{S_{1,2}}\sim \text{TeV}$. Where, $\mu_S^2$ in Eq.~\ref{DM_mass,VLQ_mass} is defined as negative to cancel out the large contribution from $F_a^2$. Fine-tuning of this type is a common feature of these axion models~\cite{Giddings:1988cx,Coleman:1988tj,Kamionkowski:1992mf,Ghigna:1992iv, Dasgupta:2013cwa}. 
The $\epsilon_s$ term of Equation~\ref{DM_mass,VLQ_mass} is the only term that causes the mass splitting between the scalar components, and $\epsilon_s$ is the dependent parameter when $M_{S_{1}}$ and the mass difference $\Delta M=M_{S_{2}}-M_{S_{1}}$ are assumed to be the independent parameters. In order to achieve $\sim \mathcal{O} ( \text{GeV})$ mass gap between $S_1$ and $S_2$, $\epsilon_s$ needs to be extremely small because $F_a$ is large. Nevertheless, the justification for a small value of $\epsilon_s$ can be found in naturalness arguments. As $\epsilon_s$ approaches zero, an additional $U(1)$ symmetry emerges in the theory, with the corresponding charges of $S,\Psi_L,\Psi_R\sim 1 $ independent of $U(1)_{PQ}$. Next, we briefly outline different constraints for this model.

\paragraph{Constraints:} 
Theoretical bounds, different experimental data, and cosmological observations severely restrict the parameter space of the extended KSVZ model. We quickly outline each of these constraints before establishing benchmark points that yield the correct relic density while accommodating all the other constraints. 

The scalar potential should be bounded from below, and the perturbativity demands all the $\lambda$ in Equation~\ref{potential_5} should be less than $4\pi$, and $|f_{i}|<\sqrt{4\pi}$. Being two component DM, the total relic density comprises both the scalar and axion DM.
\begin{equation}
\Omega_{\text{T}}h^2=~\Omega_{a}h^2+\Omega_{S_1}h^2
\label{totalRelic_5}
\end{equation}
The parameter spaces ought to match the Planck measurements~\cite{Aghanim:2018eyx} of the observed abundance of DM relics. 
\begin{equation}
\Omega_{\text{DM}}h^2=~0.120\pm 0.001.
\label{planck_5}
\end{equation}
Axions can be created non-thermally due to the misalignment mechanism, and axion relic density is as follows~\cite{Dasgupta:2013cwa, Chatterjee:2018mac, Bae:2008ue}.
\begin{equation}
\Omega _a h^2 \simeq 0.18 \hspace{1mm} \theta_a^2 \hspace{1mm} \bigg(\frac{F_a}{10^{12} \text{GeV}}\bigg)^{1.19} \, 
\label{axion_relic_5}
\end{equation} 
Where $\theta_a=a_i/F_a$ is the axion's initial misalignment angle. The axion field stays fixed at its initial value ($a_i$) until the Hubble expansion rate drops below the axion mass. The range of the axion decay constant is $10^{10}~\text{GeV}\leq F_a \leq 10^{12}~\text{GeV}$. The lower bound comes from supernova cooling data \cite{Raffelt:1987yt}, but the upper bound comes from axion overproduction. Equation~\ref{axion_relic_5} shows that axion alone may provide $100\%$ of the DM relic density if $F_a$ is properly calibrated. As previously indicated, we added the additional scalar to prevent this kind of fine-tuning. As a result, the axion maintains its underabundance, and the axion's relic density fulfils the Planck limit when combined with the scalar. Hence, for demonstrative purposes, we used $F_a=10^{11}~\text{GeV}$ and $\theta_a=1.0$, which gives the axion relic density $\Omega _a h^2 \simeq 0.012$ (approximately $10\%$ of the total observed relic), and the scalar DM delivers the rest. We also verified that even when a more substantial contribution is coming from the axion, such as the axion accounting for $50\%$ (or more) of the total dark matter relic density, there are no significant changes in the qualitative features of the allowed parameter spaces and the LHC phenomenology.
\par
%
%

Because of the mixing between the Higgs boson and $\sigma (x)$, the strength of the LHC di-photon channel turns out to be
\begin{equation}
\mu_{\gamma \gamma} = \cos^2\theta \frac{BR_{h \to \gamma \gamma}}{BR_{h \to \gamma \gamma}^{\text{SM}}}~.
\end{equation}
 Here, $\theta$ is the mixing angle between the Higgs boson and $\sigma(x)$, and LHC limits this mixing angle to $|\sin\theta |< 0.36$ \cite{Robens:2016xkb}. Higgs can decay into a pair of DM if $m_{S_1}\leq~\frac{m_h}{2}$, contributing to the invisible Higgs decay branching ratio. In our study, we assign the $hS_1S_1$ coupling to a tiny value, $\lambda_{SH}=0.01$, so that the Higgs invisible decay branching-ratio constraint is satisfied. We mainly focus on the parameter space where $m_{S_1}>~\frac{m_h}{2}$, as the majority of allowed parameters reside in this region. In this region, $\lambda_{SH}$ is unconstrained by the Higgs invisible decay branching ratio due to the kinematic forbiddance of Higgs decay into a pair of dark matter. Notably, our collider and dark matter phenomenology remain independent of the $\lambda_{SH}$ value, as the VLQ pair production cross-section at the LHC is unaffected by $\lambda_{SH}$, and dark matter predominantly annihilates into the $t\bar{u},\bar{t}u$ or $t\bar{t}$ final state through t-channel VLQ-mediated processes. The allowed parameter spaces that we will see in Fig. \ref{scan_plot_5} are also permissible for $\lambda_{SH}< 0.01$ from direct detection constraint. However, when setting a larger value for $\lambda_{SH}$, one must check the direct detection bound.

\textbf{Collider constraints:} The reinterpreted LEPII squark search results \cite{Giacchino:2015hvk, OPAL:2002bdl} exclude masses of VLQ up to 100 GeV. Please follow the brown region in Fig.~\ref{scan_plot_5}. When the mass difference between VLQ and DM, $\Delta M_{\Psi S_1}< m_t$, VLQ can decay into DM with light quarks; hence, the ATLAS search \cite{Marjanovic:2014eca} for multijet plus missing transverse momentum can further limit this scenario. The reinterpreted result \cite{Giacchino:2015hvk} of the ATLAS search \cite{Marjanovic:2014eca} for $n$-jet (2-6 jets) plus missing energy at 8 TeV center-of-mass energy with 20.3 $fb^{-1}$ integrated luminosity, excludes the purple region in Fig.~\ref{scan_plot_5}, considering VLQ decay into DM and light quark with $100\%$ branching ratio. The noteworthy feature is that, unlike in ref. \cite{Ghosh:2022rta}, when $\Delta M_{\Psi S_1}< m_t$, minor parameter spaces exist in this situation (below the red dotted line in Fig.~\ref{scan_plot_5}). 

Flavor constraints can appear through the interaction term (the first term in Equation~\ref{Lag.VLQ_5}) contributing to the $D^0-\bar{D}^0$ oscillation \cite{Garny:2014waa}. The box diagram $u\bar{c}\rightarrow \bar{u} c$ through VLQ and the scalars ($S_1, S_2$) at the loop are the Feynman diagrams contributing to this oscillation. In the current setup, the effective operator contributing to this mixing is as follows. 
\begin{equation}
\mathcal{L_\text{eff}}=\frac{\tilde{z}}{M_{\Psi}^2}\bar{u}_R^\alpha\gamma^\mu c_R^\alpha \bar{u}_R^\beta \gamma_\mu c_R^\beta,
\label{Effec_Lag_5}
\end{equation}
where $\tilde{z}=-\frac{f_u^2f_c^2}{96\pi^2}[g_{\psi}(M_{S_1}^2/M_{\Psi}^2)+g_{\psi}(M_{S_2}^2/M_{\Psi}^2)-2g_{\psi}(M_{S_1}M_{S_2}/M_{\Psi}^2)]$. The $g_{\Psi}(x)$ expression may be found in Reference~\cite{Gedalia:2009kh}. The measurement of the D-meson mass splitting yields the restriction \cite{Gedalia:2009kh, Garny:2014waa}
\begin{equation}
|\tilde{z}|\lesssim 5.7\times10^{-7}(M_{\Psi}/\text{TeV})^2~.
\label{Flavour_5}
\end{equation}

The correct relic density is achieved through $S_1 S_1 \rightarrow t\bar{q},~\bar{t}q$ ($q=u,c$) or $S_1 S_1 \rightarrow t\bar{t}$ annihilation processes in parameter spaces where $S_1$ and VLQ are not degenerate and apart from the Higgs resonance.

The parameter spaces that give correct relic density and are also allowed from DD experiments for the democratic choice of all equal coupling strengths $f_u=f_c=f_t$ are practically forbidden by the preceding flavor restriction (Equation \ref{Flavour_5}) \footnote{The flavor constraint disfavors equal coupling strength $f_u=f_c=f_t$, even for larger axion relic density. For instance, when considering $F_a=10^{11.6}~\text{GeV}$ and $\theta_a=1.0$, the axion contribution constitutes $50\%$ of the total observed relic, yet the flavor constraint still disfavors it.}. In the case of $M_\Psi$ around or above a few tens of TeV, we find that although all the parameter spaces are allowed by flavor constraints, they fail to provide the correct relic density (overabundant by a few orders) and are hence uninteresting. 
%
Instead of making two of these three couplings negligibly small, another interesting scenario emerges if we choose one of $f_u$ or $f_c$ vanishingly small (or zero) while the other remains democratically as large as $f_t$. As a result, all parameter spaces that generate correct relic density are concurrently allowed by the flavor constraint.
Hence, $\{ f_c = f_t,~ f_u=0 \}$ or $\{ f_u = f_t,~ f_c=0 \}$ are both plausible choices. When $f_u$ is nonzero, the production cross-section of the VLQ pair at the LHC differs from the scenario where $f_c$ is nonzero, primarily due to parton distribution functions. Consequently, Yukawa coupling dependence on the VLQ pair production at the LHC is more pronounced for the first-generation Yukawa coupling ($ f_u\neq 0, f_c=0 $) compared to the second-generation Yukawa interaction ($f_u= 0, f_c\neq 0 $). However, we opted for $f_u = f_t\neq 0,~ f_c=0 $ because it generates more constrained parameter spaces from the direct detection experimental upper bound. The nucleon comprises the light quarks and the gluon; therefore, $f_u\neq 0$ results in a tree-level DD scattering diagram, $S_1 u (\bar{u})\rightarrow S_1 u (\bar{u})$, via VLQ exchange (see direct detection diagrams \ref{directDetection_5}). 


\section{Dark Matter Phenomenology}
\label{DMphenomenology_5}

This section examines the dark matter phenomenology due to the scalar component. Before we get into the details, let's look at the relevant free parameters. Since axion couplings with scalar DM or SM particles are inversely proportional to $F_a$, such couplings are severely suppressed and have practically no role in scalar DM phenomenology. The large mass and decoupling of $\sigma(x)$ from the remaining particle spectrum ensure that it has no impact on DM phenomenology. The relevant parameters are $\{M_{\Psi}, M_{S_1}, \Delta M, f \}$. As previously stated, one way to bypass the prohibitory flavor constraint is by setting $f_{c}=0$. 

\paragraph{Relic density of DM:}
\begin{figure}[tb!]
\centering
\includegraphics[scale=0.5]{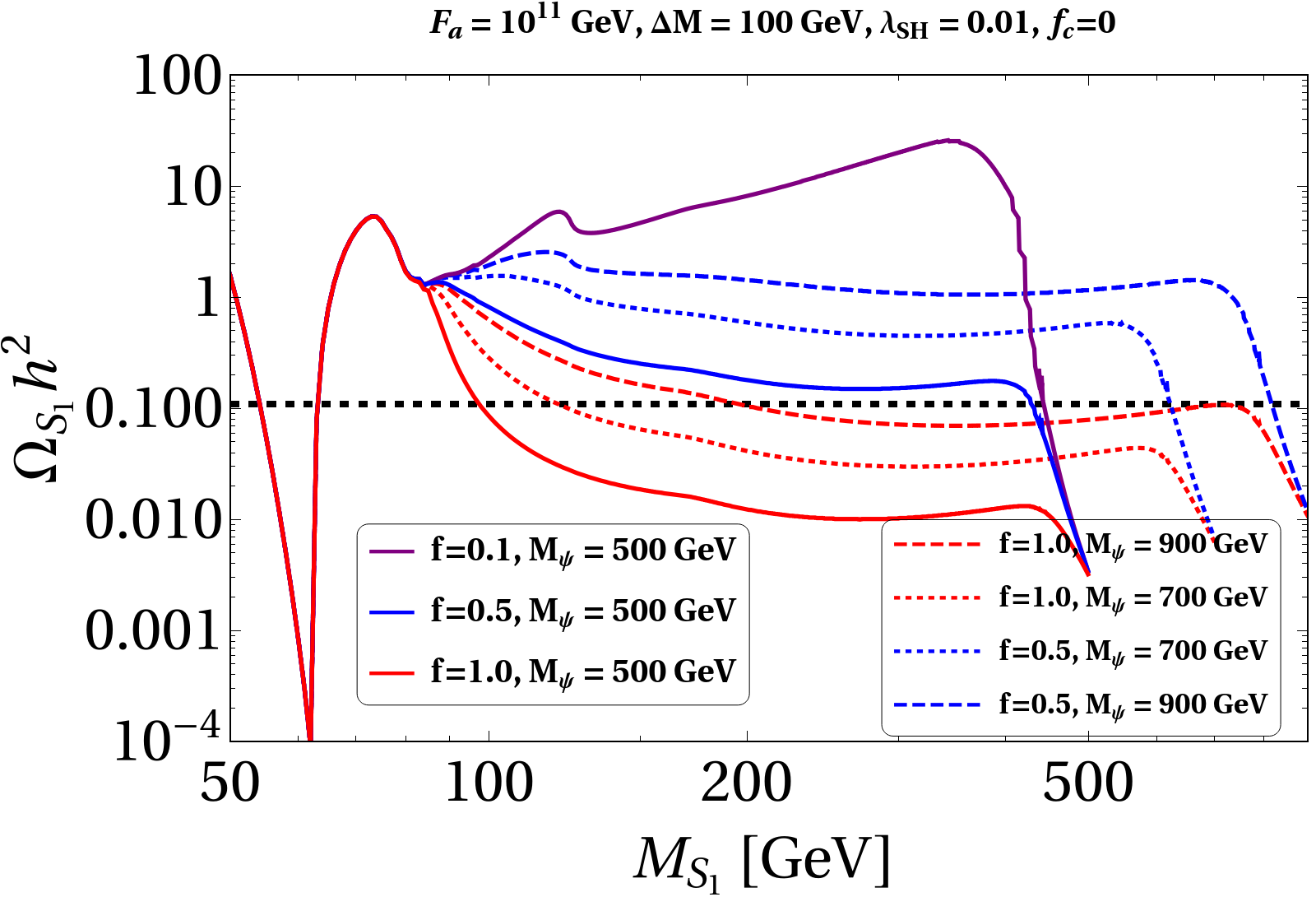}
\caption{Variations of the scalar DM relic density with its mass ($M_{S_1}$) for different values of $f$ and $M_{\Psi}$ are shown. Here, we fix $F_a=10^{11}~\text{GeV},~\lambda_{SH}=0.01$, $f_c=0$, and $\Delta M=100$ GeV. The Black dashed line corresponds to $0.120-\Omega_ah^2$. 
}
\label{relic_plot_5}
\end{figure}
%
\begin{figure}[tb!]
\centering
\includegraphics[scale=0.35]{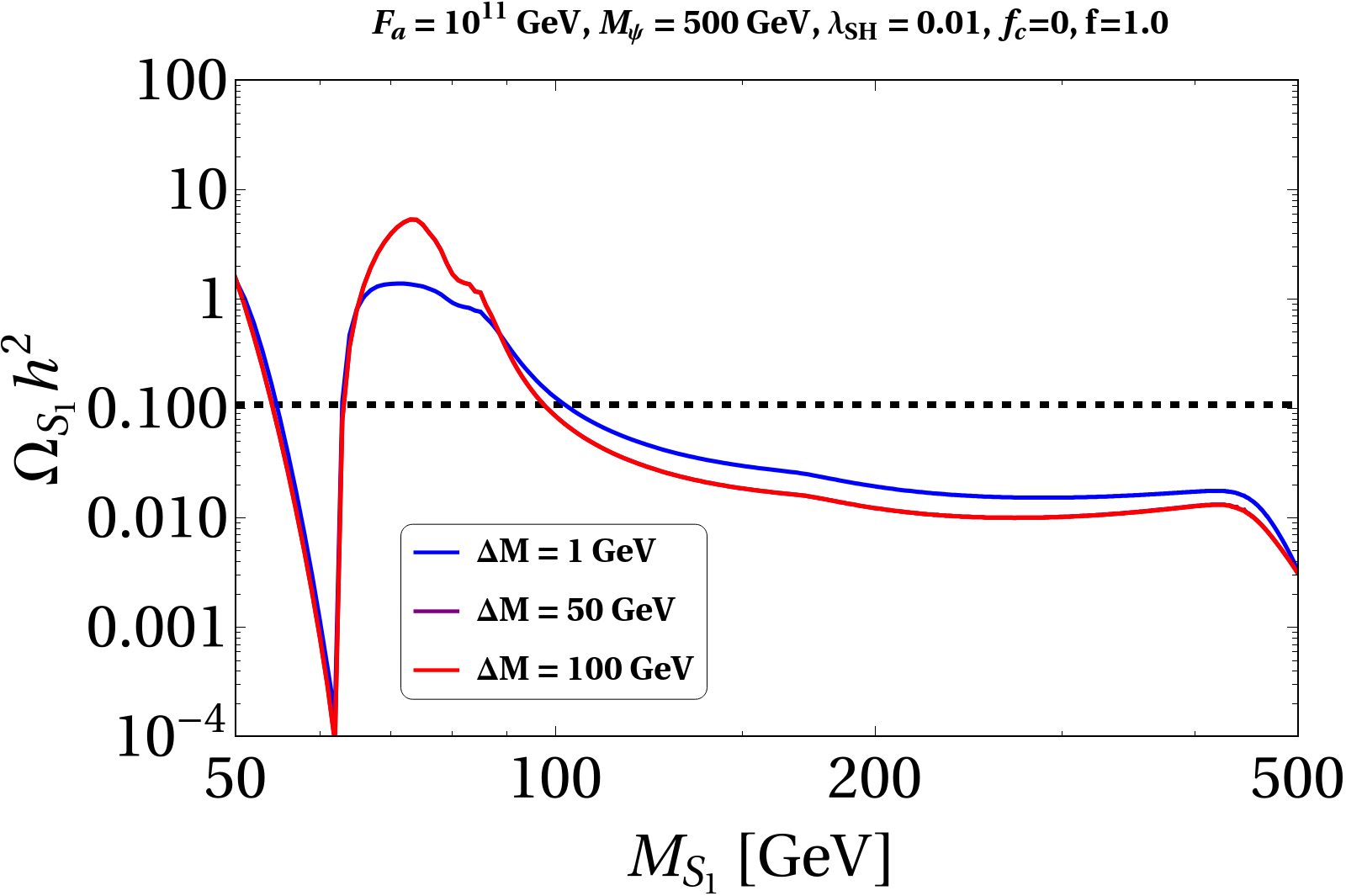}
\includegraphics[scale=0.35]{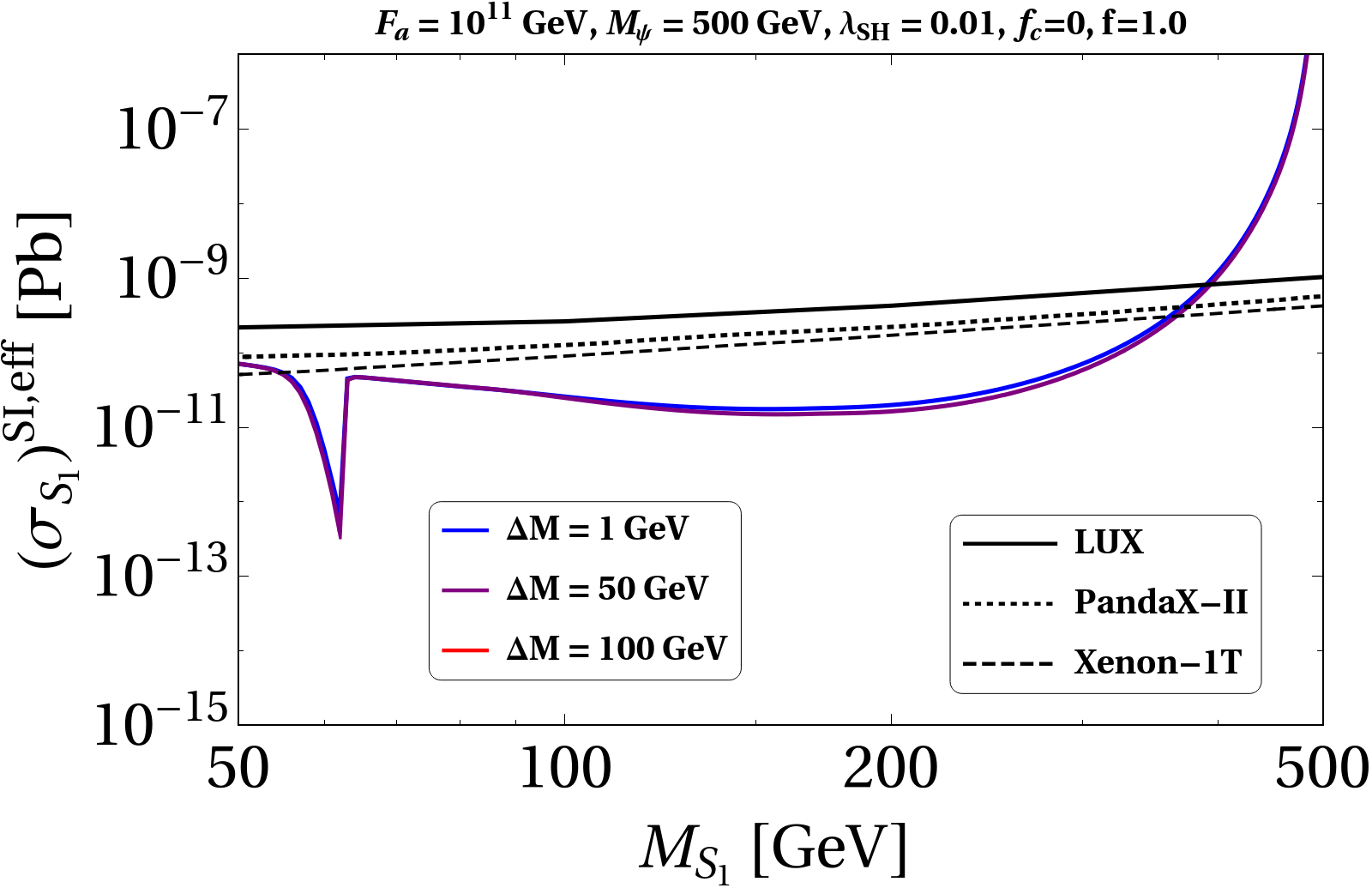}
\caption{Variations of the scalar DM relic density (left panel) and effective direct detection cross-section [Eq.~\ref{Eq_eff_DD_5}] (right panel) with its mass ($M_{S_1}$) for different values of $\Delta M=M_{S_2}-M_{S_1}$ are shown. Here, we fix $F_a=10^{11}~\text{GeV},~\lambda_{SH}=0.01$, $f_c=0$, $f=1$ and $ M_\Psi=500$ GeV. The Black dashed line in the left plot corresponds to $0.120-\Omega_ah^2$. The red and purple are identical in the left plot.
}
\label{relic_DD_varryDelM}
\end{figure}
%

To estimate the component of relic density offered by scalar DM, we solve the Boltzmann equation using \texttt{micrOMEGAs} -v5 \cite{Belanger:2018ccd}. We first construct our model in Feynrules~\cite{Alloul:2013bka}. The variation of the scalar DM relic density with its mass is displayed in Figure~\ref{relic_plot_5} while we fix $F_a=10^{11}~\text{GeV},~\lambda_{SH}=0.01$, $f_c=0$, and $\Delta M=100$ GeV. We present three solid lines for three distinct values of democratic coupling $f=0.1,~0.5,$ and $1.0$ for the 500 GeV mass of the VLQ. In these variation curves, the first sharp dip ensues due to the Higgs resonance, in which pair of DM annihilate into the SM particles through the resonant Higgs boson when $M_{S_1}\sim \frac{m_h}{2}$, while the second dip occurs when $M_{S_1}\sim M_{W}$, in which pair of $S_1$ annihilate into a $W$ boson pair through s-channel Higgs-mediated diagram, see Figure~\ref{annihilation_5}. 
 
For $f=0.1$ (solid purple line), after the second dip, the relic density increases along with the increase in the DM mass, and a third dip is observed at $M_{S_1}=m_h$. The pair of $S_1$ begin to annihilate into the Higgs bosons via contact interaction, Higgs-mediated s-channel, and $S_1$-mediated t-channel diagrams (Figure~\ref{annihilation_5}) and produce the third dip. Ultimately, when the mass difference between VLQ and DM becomes smaller, the impact of DM co-annihilation with the VLQ and annihilation of the VLQ pair into gluons becomes apparent, and a final decline in DM relic density is observed. 
Further increasing $f$ (0.5 with solid blue line and 1.0 with solid red line) reveals that relic density declines just after the second dip due to the significant contribution of $S_1 S_1 \rightarrow t\bar{u},~\bar{t}u$ annihilation channels via the VLQ-exchange t-channel processes. The correct relic density is achieved for $f=1.0$ when DM mass is around $96$ GeV.

Blue (red) dotted and dashed lines correspond to the same values of $f$ as in solid lines, except with a heavier choice of mediator ${\Psi}$. Because the annihilation cross section decreases as propagator mass increases, and relic density is inversely proportional to the annihilation cross section, the dotted and dashed lines move to higher relic density than the solid line. 
One clearly follows from these variations that significant parameter space for heavier dark matter masses can open up for different choices of these parameters (over and above the typical Higgs portal).
Interestingly, in the case of a pure scalar singlet DM scenario, the DM does not satisfy the correct relic density for $\lambda_{SH}=0.01$. However, the interaction of the DM with the SM top quark in the present model affords many parameter spaces that satisfy the Planck limit.

Dominant parameter space provides correct relic density because of the annihilation processes $S_1 S_1 \rightarrow t \bar{u}, t \bar{t}$ mediated by $\psi$. If the mass of the VLQ, $M_{\psi} \sim \mathcal{O} (F_a)$  (or even around 10 TeV), the annihilation cross section becomes significantly small. Consequently, the scalar DM exceeds the dark matter relic density by several orders of magnitude. In such a scenario, the model loses its attractiveness and appeal.
%

Fig. \ref{relic_DD_varryDelM} displays the variations of scalar dark matter relic density (left panel) and effective direct detection cross-section (right panel) with its mass ($M_{S_1}$) for different values of $\Delta M=M_{S_2}-M_{S_1}$. In the left plot, we observe that the variations in relic density are identical for $\Delta M=50,100$ GeV, distinct from $\Delta M=1$ GeV. This discrepancy arises because, with a large mass difference between dark matter (DM) and its heavier counterpart $S_2$, there is no coannihilation between them. Conversely, when they are nearly degenerate, coannihilation occurs, leading to changes in relic density. In the right plot, the spin-independent direct detection cross-sections are identical. This uniformity arises from the fact that $S_2$ has no role in the elastic scattering between dark matter and nucleon. As a result, the dark matter phenomenology remains unaltered for any  $\Delta M$ values as long as they are not nearly degenerate. Consequently, our analysis set $\Delta M=100$ GeV as a representative value.
\paragraph{Direct and indirect detection of DM:}
\begin{figure}[tb!]
\centering
\includegraphics[scale=0.36]{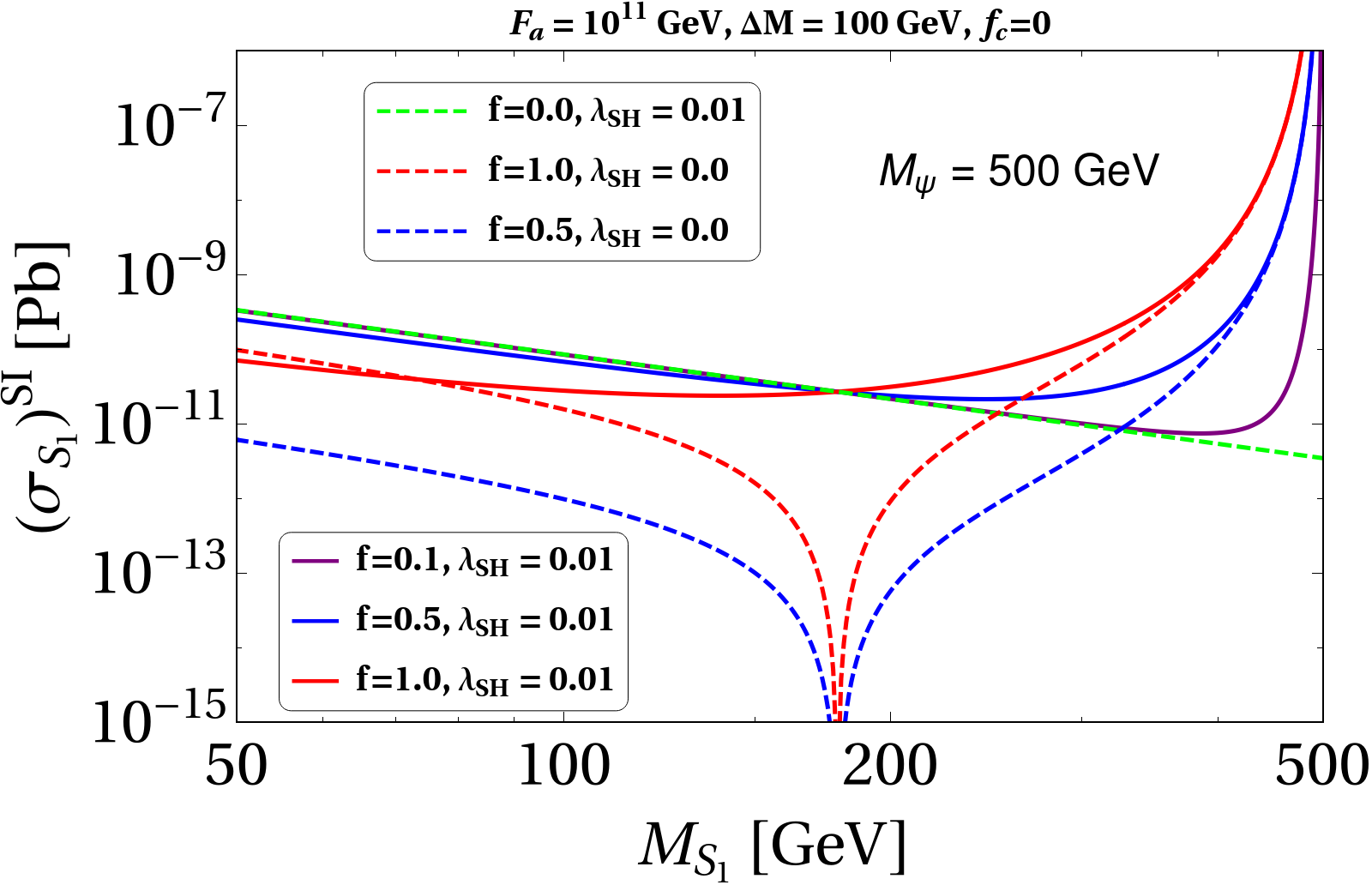}
\includegraphics[scale=0.36]{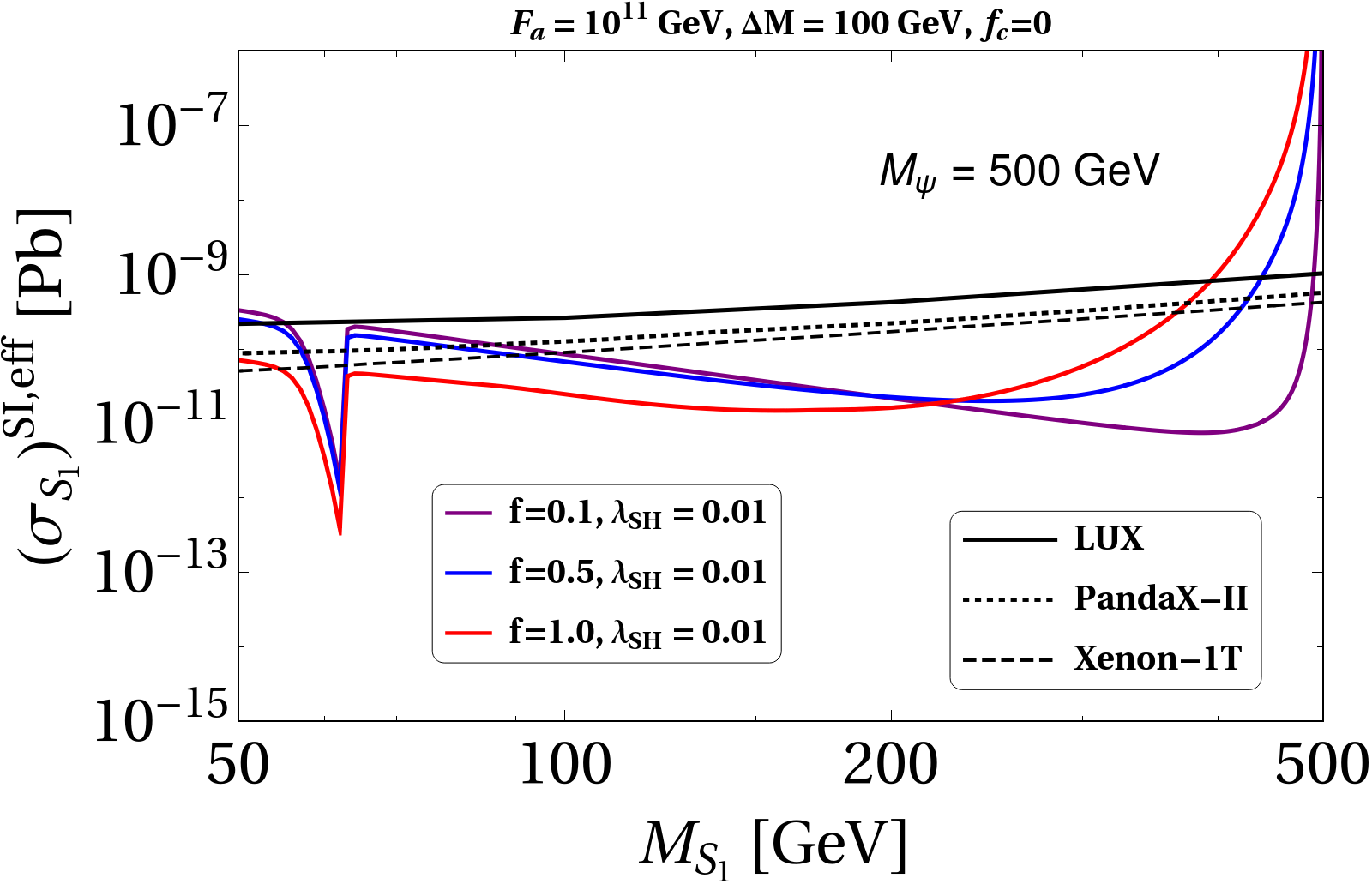}
\caption{\textbf{Left panel:} Spin-independent cross section of scattering between scalar DM and nucleon as a function of dark matter mass $M_{S_1}$. We set $f = 0, \lambda_{SH}=0.01$ (green dashed line), and $\lambda_{SH}=0$ for the blue-dashed ($f = 0.5$) and red-dashed ($f = 1.0$) lines to illustrate the individual contributions. \textbf{Right panel:} Effective spin-independent scattering cross section (Equation \ref{Eq_eff_DD_5}) vs dark matter mass. All of the plots on the left and right are for $M_{\Psi}=500$ GeV. The solid colors purple, blue, and red in both panels stand for $f=0.1,~0.5,$ and $1.0$, respectively, and $\lambda_{SH}=0.01$ for all solid lines. 
}
\label{DD_plots_5}
\end{figure}
WIMPs may also scatter off nuclei, depositing energy that can be detected by detectors like LUX \cite{Akerib:2016vxi}, PandaX-II \cite{Tan:2016zwf, Cui:2017nnn}, and XEXON1T \cite{Aprile:2018dbl}. These experiments can set strong constraints on the scattering cross section and DM mass. All Direct detection channels and the square amplitude of these diagrams are shown in Appendix \ref{appen_5}. For demonstration purposes, we present a spin-independent direct detection cross section of $S_1$ with its mass shown in the left panel of Figure~\ref{DD_plots_5}. All solid lines correspond to $\lambda_{SH}=0.01$ but for different values of $f=$ 0.1 (solid purple), 0.5 (solid blue), and 1.0 (solid red). Because $f$ and $\lambda_{SH}$ are both non-zero, the Higgs-mediated and VLQ-mediated channels and their interference diagrams contribute. It is instructive to note how individual channel contributes. One can first set $f=0, ~\lambda_{SH}=0.01$ (dashed green line) so only the Higgs-mediated channel contributes. Subsequently, setting $\lambda_{SH}=0$ for two choices of $f=1.0~(0.5)$ in the dashed-red (blue) line demonstrates the contribution from pure VLQ-mediated s and t-channels and their interference diagram. The Higgs-mediated diagram does not contribute here.
 
The amplitude square of the Higgs-mediated diagram does not rely on the mass of $S_1$ (Equation~\ref{DD_hh}); nevertheless, the cross section of the dashed green line decreases with the DM mass, which comes from the phase space part of the integral. We see dashed red and blue lines strongly depend on the $M_{S_1}$ since the amplitude square of the VLQ mediated s and t- channels and their interference explicitly depends on $M_{S_1}$ (see Equations \ref{DD_ss}- \ref{Eq_num_5}), and the cross section is minimum when $M_{S_1}=\dfrac{M_{\Psi}}{2\sqrt{2}}$. When comparing dashed-green (only Higgs-mediated channel contributes), dashed-red (only VLQ-mediated channels contribute), and solid-red (total cross section) lines, one can witness a substantial negative (positive) interference between Higgs and VLQ-mediated diagrams when $M_{S_1} < \dfrac{M_{\Psi}}{2\sqrt{2}}$ ($M_{S_1} > \dfrac{M_{\Psi}}{2\sqrt{2}}$).
Finally, when DM mass is large, we see a sharp rise because of the on-shell production of VLQ (Figure \ref{DD_s}).
 
In a two-component DM scenario, the direct detection cross section of the scalar DM should be rescaled as 
\begin{equation}
\sigma_{S_1}^{\text{SI,\text{eff}}}=\Bigl(\frac{\Omega_{S_1}h^2}{\Omega_{\text{T}}h^2}\Bigr)~\sigma_{S_1}^{\text{SI}},
\label{Eq_eff_DD_5}
\end{equation}
where $\Omega_{\text{T}}h^2$ is given in Equation \ref{totalRelic_5}. The spin-independent effective direct detection cross section for three distinct values of $f=0.1,~0.5,$ and $1.0$ are presented in the right panel of Figure \ref{DD_plots_5} (same as solid lines in the left panel). The black lines show the experimental upper bounds. A dip around $M_{S_1}\sim \frac{m_h}{2}$ is found because of rescaling, as given in Equation \ref{Eq_eff_DD_5}. Here, one notices that the DD experiments disallow the region with a considerable mass difference between the VLQ and DM for significantly large $f$ values. For example, the regions when $M_{S_1}>390$ GeV for $M_\Psi=500$ GeV and $f=1$ (solid red line) are disallowed by DD. Additionally, Figures \ref{relic_plot_5} and \ref{DD_plots_5}, for $f=1$, suggest that the parameter spaces that offer correct relic density are likewise allowed from the DD.  
 
WIMPs may self-annihilate, emitting a significant amount of gamma and cosmic rays while looking at the dense DM regions at the galactic centre. Indirect detection experiments like PAMELA~\cite{Kohri:2009yn}, Fermi-LAT~\cite{Eiteneuer:2017hoh}, MAGIC~\cite{Ahnen:2016qkx} etc. can constrain model parameter spaces substantially. Because this is a two-component DM scenario, the scalar DM's indirect detection cross section should also be rescaled as,
\begin{equation}
\sigma^{ID}_{S_1, \text{eff}}=\Bigl( \frac{\Omega_{S_1}h^2}{\Omega_{\text{T}}h^2} \Bigr)^2~\sigma^{ID}_{S_1}~.
\label{Eq_eff_ID_5}
\end{equation}
 We note that most indirect detection experiments enable our model because the necessity for the axion keeps $S_1$ under abundance, which lowers the indirect detection constraints since it depends on the fractional scalar DM relic density squared. For very small $\Delta M$ $=(M_{S_2}-M_{S_1}$), $S_1$ and $S_2$ are nearly degenerate and can open an additional channel where $S_1$ and $S_2$ co-annihilate into top anti-top pair via VLQ and contribute to the indirect detection cross section, which may be disallowed by antiproton cosmic ray data~\cite{Colucci:2018vxz}, so we set $\Delta M=100$ GeV throughout our analysis, ensuring that no co-annihilation channel exists. 

\begin{figure}[tb!]
\centering
\includegraphics[scale=0.7]{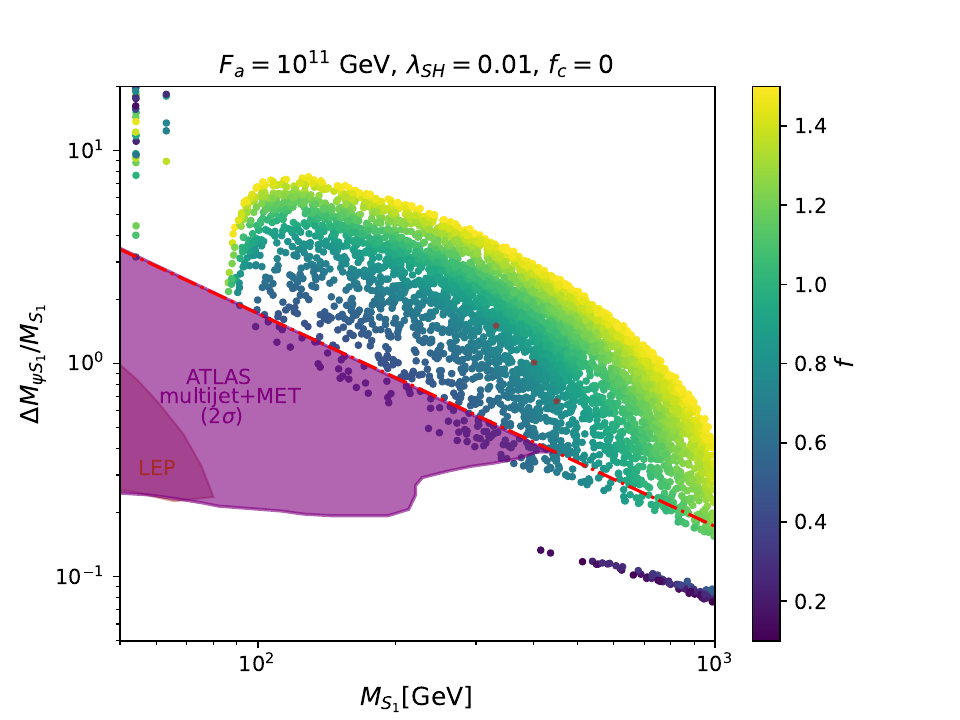}
\caption{On the $\frac{\Delta M_{\Psi S_1}}{M_{S_1}} - ~M_{S_1}$ plane, the parameter spaces that satisfy the measured DM abundance, permitted by the direct search experiments, and comply with other restrictions as stated in the text are displayed. The color coding is done with respect to $f$, with $f$ varying from 0.1 to 1.5. Here, we fix $F_a=10^{11}~\text{GeV},~\lambda_{SH}=0.01$, $\Delta M =100$ GeV and $f_c=0$. The red dash-dot line corresponds to $\Delta M_{\Psi S_1}=m_t$ (mass of top-quark). Three brown stars show three benchmark points in Tab.\ref{parameters_5}. The $2\sigma$ exclusion regions from the LEP and ATLAS (multijet $+$ MET) analyses are represented by brown and purple regions, respectively.  
}
\label{scan_plot_5}
\end{figure}

It is interesting to note that estimating the relic density contribution of any individual dark matter component from direct or indirect searches or collider experiments is notably challenging~\footnote{These difficulties arise due to multiple factors involving dark matter evolution and the limitations  of traditional measurement techniques. In a multi-component dark matter scenario, the possibility of finding one such candidate from the experiments is significantly weaker, as each component only contributes fractionally to the total dark matter relic density (as described in Eqs. \ref{Eq_eff_DD_5} and \ref{Eq_eff_ID_5}). In fact, with gradually strong constraints already in place, such undetectability is one of the primary motivations for the recent euphoric development of different multi-component dark matter scenarios \cite{Konar:2009qr} or efforts to develop various unconventional production mechanisms of dark matter, such as freeze-in \cite{Hall:2009bx, Ghosh:2021wrk,Chakrabarty:2022bcn}.
Even if we could pinpoint one such dark matter candidate, the exact contribution in the calculated relic density would crucially depend on several factors like dark matter production mechanism (e.g. Freeze-out, Freeze-in, FIMP, SIMP, etc.), the cosmological evolution (whether standard or non-standard \cite{DEramo:2017gpl,Das:2023owa} dominated background) of the universe during the same. Another critical factor is that the dominant processes that contribute to the relic density might not be the channels that contribute to direct or indirect detection interactions. A richer structure in the dark sector, the corresponding mass spectrum and their interrelation can make the scenario more complex, with significant contributions through annihilation, co-annihilation, resonance and interference.}. However, one may still seek to make specific projections about their relative contributions confined within a simplified model like ours, assuming the standard freeze-out WIMP production mechanism. Note that, in our WIMP-axion multi-component scenario, the processes that contribute to the WIMP relic density and direct detection primarily arise due to a (democratic) Yukawa interaction between $S_1$, t(u) and VLQ, and required parameters being mass of VLQ ($M_\Psi$), the mass of DM ($M_{S_1}$) and the Yukawa coupling $f$. For demonstration purposes, in Figure \ref{scan_plot_5}, we showed relic satisfying points for different $M_\Psi$, $M_{S_1}$ and $f$ values for a sample contribution of axion dark matter of approximately 10\% of the total observed relic.
Constraints on these parameters can, in turn, provide a bound over the contribution of WIMP dark matter~\footnote{For example, see \cite{Konar:2009ae}, where constraints over masses and models are estimated model-independently, given that the same interaction generated the observed relic.}. Now, we will briefly discuss the axion detection techniques.

The axion, which appears in the extension of the Standard Model, provides a dual solution to two fundamental problems: serving as a dark matter candidate and addressing the strong CP problem. Current and upcoming experimental searches for axions have the potential to offer crucial insights into their existence. However, most of these searches place limits on the axion mass under the assumption that the axion's energy density accounts for the entirety of the dark matter halo's density, $\rho_{\text{DM}}^{\text{halo}}\approx 0.45~ GeV~cm^{-3}$.

Numerous axion experiments are underway globally, and we will briefly discuss some of them. Many experiments exploring axion dark matter rely on its interaction with the electromagnetic field, described in Equation \ref{Eq.axion_search}, and known as `haloscopes.' 
\begin{equation}
\mathcal{L}_{\text{eff}} \supset  \dfrac{1}{4} g_{a\gamma\gamma}~ a(x)~ \tilde{F}_{\mu \nu} F^{\mu \nu}= g_{a\gamma\gamma}~ a(x)~ \textbf{E}~\cdot~\textbf{B}
\label{Eq.axion_search}
\end{equation} 
$F^{\mu \nu}$ ($\tilde{F}_{\mu \nu}$) denotes the (dual) electromagnetic field strength tensor, with $\textbf{E}$ and $\textbf{B}$ representing the electric and magnetic fields, respectively. The axion photon coupling, $g_{a\gamma\gamma}$, is proportional to the electromagnetic fine-structure constant and axion mass \cite{Ringwald:2024uds}. Different haloscope experiments target different mass ranges of the axion. For example, ADMX \cite{Stern:2016bbw}, BabyIAXO \cite{Ahyoune:2023gfw}, and FLASH \cite{Alesini:2023qed} target the axion mass range $0.5~\mu eV \lesssim m_a \lesssim 100~\mu eV$. In these experiments, a microwave cavity is placed within a strong magnetic field to detect axions through their conversion into photons.  The output power amplifies if the axion's mass matches the cavity's resonance frequency.  Since the exact mass of the axion is unknown, the axion mass range can be explored by tuning the cavity's resonance frequency. A proposed experiment, MADMAX \cite{Caldwell:2016dcw}, uses dielectric haloscopes to cover a higher mass range over $100~\mu eV$. Another avenue to explore axions is to search solar axions. The CAST experiment \cite{CAST:2017uph} at CERN looks for solar axions generated from the interaction between photons and the Coulomb fields of nuclei within the solar plasma.  Although the axion's relative contribution to dark matter is still unknown, ongoing experiments are refining the range of axion masses and offering insights into its role as dark matter.

\paragraph{Parameter scan and benchmark points:}
\begin{table}[tbhp!]
\begin{center}
 \resizebox{\columnwidth}{!}{
 \begin{tabular}{|c|c|c|c|c|c|c|c|c|}
\hline\hline
 & $M_{S_1}$& $\Delta M_{\Psi S_1}$ & $\Delta M$ & $f$ & $\Omega_{S_1} h^2$ & $\sigma^{SI}_{S_1, eff}$ & Processes &  \\ 
    & (GeV)        &      (GeV)                      &         (GeV) &  &  &  (pb) & (percentage) & BR($\Psi\rightarrow t S_{1,2}$) \\%
\hline\hline
BP1 & 332 & 500 & 100 & 0.83 & 0.109 & $7.98 \times 10^{-12}$ & $ S_1 S_1 \rightarrow t \bar{u},\bar{t} u ~(60\%)$ & 0.4907 \\
 &  &  &  &  & &  & $ S_1 S_1 \rightarrow t \bar{t}~(40\%)$ & \\
\hline
BP2 & 402 & 407 & 100 & 0.82 & 0.109 & $7.65 \times 10^{-12}$ & $ S_1 S_1 \rightarrow t \bar{u},\bar{t} u ~(56\%)$ & 0.4875\\
 &  &  &  &  & &  & $ S_1 S_1 \rightarrow t \bar{t}~(44\%)$ & \\
\hline
BP3 & 450 & 300 & 100 & 0.79 & 0.107  & $1.14 \times 10^{-11}$ & $ S_1 S_1 \rightarrow t \bar{u},\bar{t} u ~(54\%)$ & 0.435\\
 &  &  &  &  & &  & $ S_1 S_1 \rightarrow t \bar{t}~(45\%)$ & \\
\hline
 \end{tabular}
 }
 \caption{A few representative benchmark points (BPs) from the scan plot are presented; these BPs satisfy the correct relic density and are permissible under all constraints. $\Omega_{S_1} h^2$ and $\sigma^{SI}_{S_1, eff}$ (Equation \ref{Eq_eff_DD_5}) are the relic density and the effective direct detection cross section of the scalar DM, $S_1$, respectively. $\Delta M_{\Psi S_1}=M_\Psi -M_{S_1}$ and $\Delta M=M_{S_2}-M_{S_1}$. Other parameters are $F_a=10^{11}~\text{GeV},~\lambda_{SH}=0.01,~\text{and}~ f_{c}=0$. The second last column shows the different processes that contribute to the relic density with the percentage contributions in the bracket. The branching fraction of VLQ decays into the top quark associated with the scalar is shown in the last column.
}\label{parameters_5}  
\end{center}
\end{table}
\begin{table}[tb!]
\begin{center}
 \resizebox{\columnwidth}{!}{
 \begin{tabular}{|c|c|c|c|c|c|}
\hline\hline
 & $\sigma_{S_{1}}^{\text{ID}}$ & $ S_1 S_1 \rightarrow t \bar{u},\bar{t} u $ & $S_1 S_1 \rightarrow t \bar{t}$  & $\sigma^{ID}_{S_1, \text{eff}}$ (in $t \bar{t}$) & $\sigma^{ID}_{exp}$  (in $t \bar{t}$)\\ 
BP    & ($cm^3/s$)  & (in $\%$)  & (in $\%$) & ($cm^3/s$) & ($cm^3/s$)  \\%
\hline\hline
BP1 & $2.30 \times 10^{-26}$ & 59.8 & 40.0 & $7.59 \times 10^{-27}$ & $2.37 \times 10^{-26}$ \\
\hline
BP2 & $2.36 \times 10^{-26}$ & 56.0 & 43.7 & $8.51 \times 10^{-27}$ &  $2.54 \times 10^{-26}$ \\
\hline
BP3 & $2.43 \times 10^{-26}$ & 54.4 & 45.4 & $8.77 \times 10^{-27}$ & $2.60 \times 10^{-26}$ \\
\hline
 \end{tabular}
 }
 \caption{Total indirect detection cross section, $\sigma_{S_{1}}^{\text{ID}}$ and the percentage of the different processes that contribute to indirect detection are presented in the third and fourth columns. Effective indirect detection cross section in the $t \bar{t}$ final state for different benchmark points is shown in the fifth column, which is defined as $(\text{\% contribution in}~ t \bar{t}~\text{final state}) \times \sigma^{ID}_{S_1, \text{eff}}$, where $\sigma^{ID}_{S_1, \text{eff}}$ is given in Equation~\ref{Eq_eff_ID_5}. The last column is the experimental upper bound in the $t\bar{t}$ final state ~\cite{Colucci:2018vxz}.
}\label{Indirec-Detection_5}  
\end{center}
\end{table}
To demonstrate the relevant parameter space that offers correct relic abundance while being allowed by direct detection and all other constraints as specified in the last section, we identify the three most important parameters, that is, the masses $M_{S_1}$, $\Delta M_{\Psi S_1}$ and the Yukawa coupling $f$. Figure~\ref{scan_plot_5} display such points on the plane of $M_{S_1}$ vs $\frac{\Delta M_{\Psi S_1}}{M_{S_1}}$, while  the Yukawa coupling $f$ is color coded, with $f$ ranging from 0.1 to 1.5. 
The red dash-dot line corresponds to $\Delta M_{\Psi S_1}=m_t$. Hence, the upper portions of this line can be investigated at the LHC with top quark on-shell production as VLQ decays into a top quark and invisible DM, while the lower area can be probed with jets $+$ MET as VLQ decays into an $u$-quark associated with DM. Points along two vertical lines at the top left region correspond to the part satisfied by Higgs resonance. 

It is enlightening to note that the lower sections of the plot, which correspond to the small mediator mass $M_{\Psi}$, typically generate an increased DD cross section; therefore, those regions are excluded from the DD bounds despite having the correct relic density.
Only a few points exist at the lower right corner when $f$ is tiny. Those regions have correct relic density because being nearly degenerate, the VLQ and DM co-annihilate and the pair of VLQ annihilates into gluons. Note that for larger $f$ values, those co-annihilation regions are ruled out by the DD experiments, as we already see in Figure \ref{DD_plots_5} (right panel, red, blue lines). Interestingly, non-perturbative effects like Sommerfeld enhancement and bound state formation can significantly affect relic density in those co-annihilation regions. Further study of this region is beyond the scope of the present discussion. Such points are challenging to probe at the LHC as the DM mass is quite large and VLQ is degenerate to the DM, so the partonic cross section of VLQ production will be small, and VLQ will emit a soft jet that is very difficult to detect. 

A few representative benchmark points (BPs) from the scan plot are listed in Table \ref{parameters_5}, which are allowed from all the constraints and provide correct relic density. The scalar DM relic density, spin-independent DD scattering cross section of $S_1$, the percentage contribution of each process to the relic density, and the branching ratio of VLQ decay into the top quark are also given. Table \ref{Indirec-Detection_5} shows the total cross section of indirect detection (ID) and the percentage contribution of the various processes to the indirect detection. The theoretical ID cross section in the final state of $t\bar{t}$ and the experimental upper bound are given in the last two columns, where we find that all of those BPs are well inside the experimental upper bound. 
%

\section{Pair production of vector-like quark at NLO+PS accuracy}
\label{nlops_5}

We implement the model Lagrangian discussed in Equation \ref{EQ. KSVZ_5} together with the interaction terms of Equations \ref{Lag.VLQ_5} and \ref{potential_5} in {\sc FeynRules} \cite{Alloul:2013bka} and employ the {\sc NLOCT} \cite{Degrande:2014vpa} package to generate {\sc UV} and {\sc $R_2$} counterterms of the virtual contribution in NLO UFO model that we finally use under the {\sc MadGraph5\_aMC@NLO} \cite{Alwall:2014hca} environment. Inside this, the real corrections are performed using the FKS subtraction method \cite{Frixione:1995ms, Frixione:1997np}. The older version of {\sc MadGraph5\_aMC@NLO} utilized the Ossola-Papadopoulos-Pittau (OPP) method \cite{Ossola:2006us} for integral reduction. However, the current version employs Tensor Integral Reduction (TIR)~\cite{Passarino:1978jh, Davydychev:1991va} procedures to handle the loop integrals.
Showering of the events is done using {\sc Pythia8}~\cite{Sjostrand:2001yu, Sjostrand:2014zea}. For leading order (LO) and next-to-leading order (NLO) event generation, we use {\sc NN23LO} and {\sc NN23NLO PDF} sets, respectively.

\begin{figure}[tb!]
\centering
\subfloat[$\mathcal{M}_{\text{LO}}^a=\mathcal{O}(\alpha_S)$] {\label{FD_only_QCD_5} \includegraphics[scale=0.32]{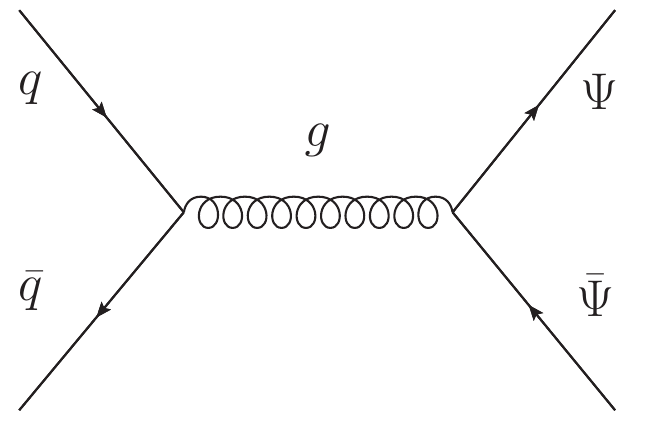}
\includegraphics[scale=0.32]{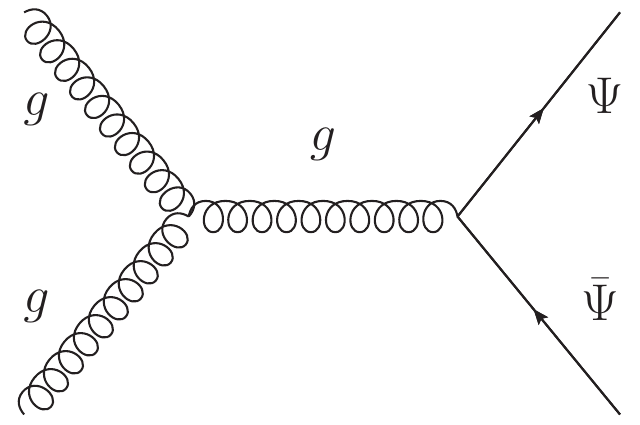}
\includegraphics[scale=0.32]{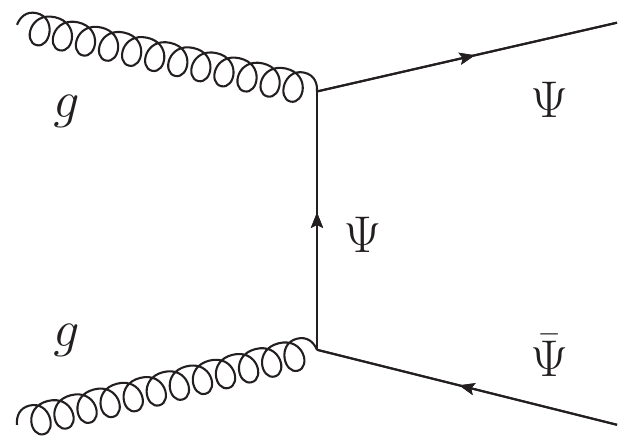}}\\
\subfloat[$\mathcal{M}_{\text{LO}}^b=\mathcal{O}(f^2)$] {\label{FD_only_NC} \includegraphics[scale=0.32]{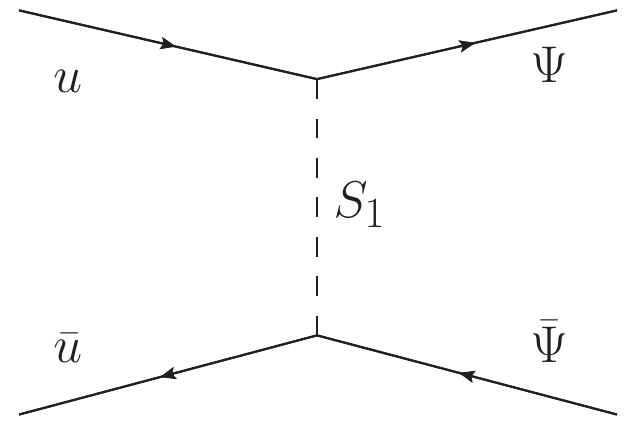}
\includegraphics[scale=0.32]{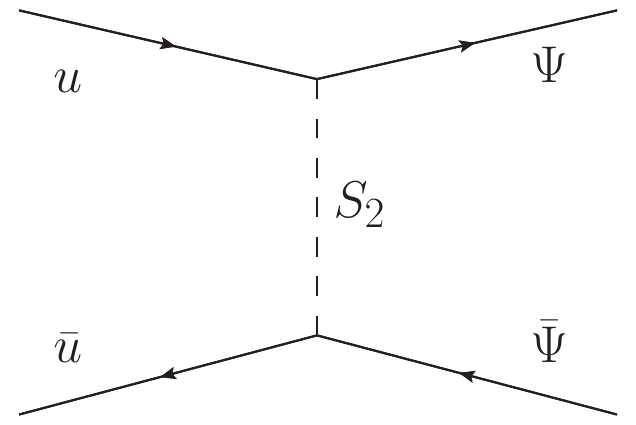}}
\caption{The Feynman diagrams for the pair production of VLQ at Leading order. 
}
\label{FD_production_5}
\end{figure}

\begin{figure}[tb!]
\centering
\includegraphics[scale=0.32]{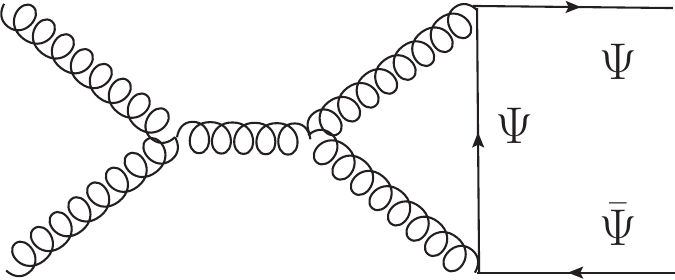}
\includegraphics[scale=0.32]{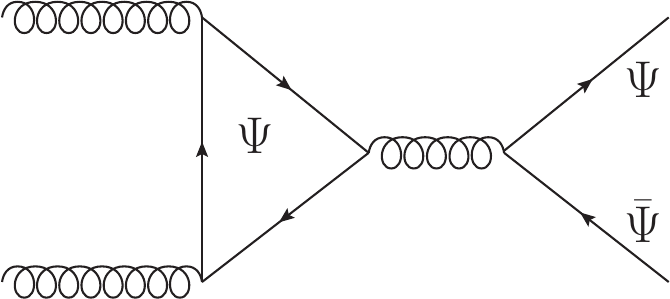}
\includegraphics[scale=0.32]{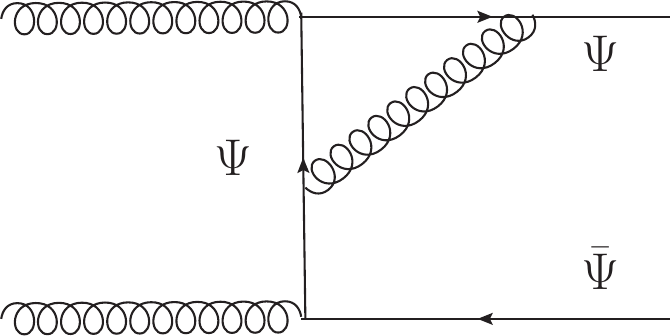}
\includegraphics[scale=0.32]{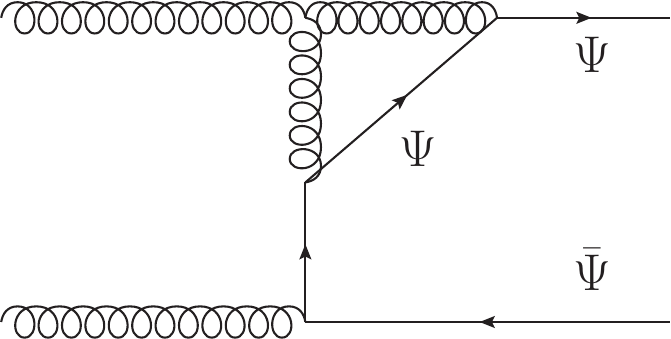}\\
\includegraphics[scale=0.32]{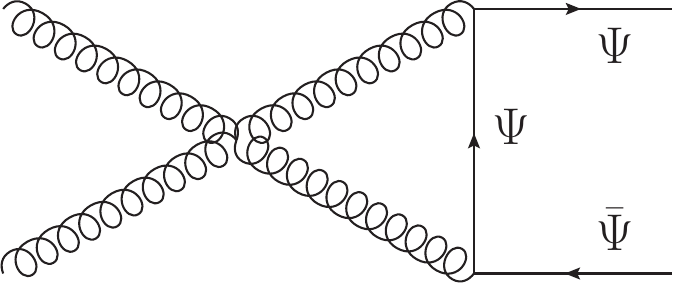}
\includegraphics[scale=0.32]{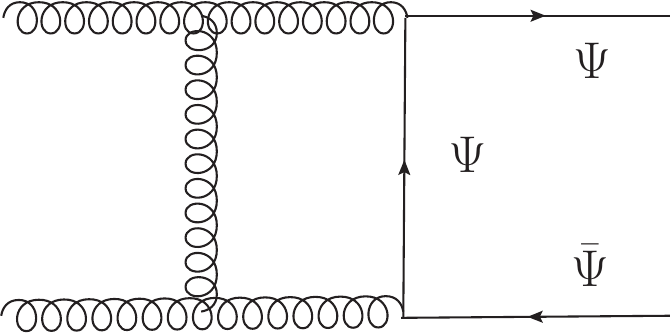}
\includegraphics[scale=0.32]{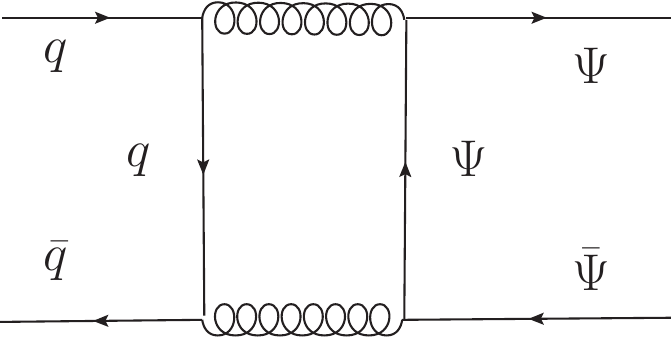}
\includegraphics[scale=0.32]{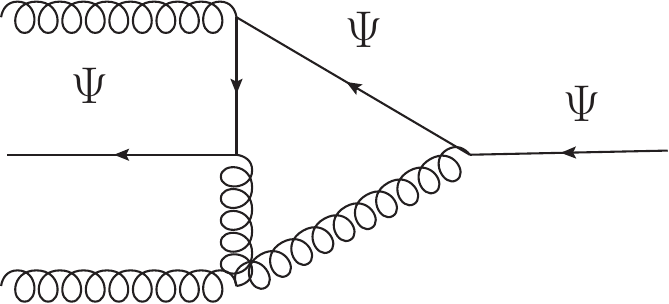}
\caption{Representative Feynman diagrams for the pair production of VLQ at NLO-QCD for the processes where the tree-level diagrams only have QCD coupling (Figure \ref{FD_only_QCD_5}). $\sigma_{\text{NLO}}^a\propto \mathcal{M}_V^\dagger \mathcal{M}_{\text{LO}}^a = \mathcal{O}(\alpha_S^3)$.}
\label{FD_production_NLO-1_5}
\end{figure}
%

All the tree-level diagrams in the pair production of VLQ at the LHC are shown in Figure~\ref{FD_production_5}. The three Feynman diagrams in the upper row depend only on the QCD coupling and are the dominant production channels. The bottom two diagrams depend on the BSM Yukawa coupling, $f$. The LO cross section has the order $\sigma_{\text{LO}}= \mathcal{O}(\alpha_S^2)+\mathcal{O}(f^4)+\mathcal{O}(f^2\alpha_S)$. The term $\mathcal{O}(f^2\alpha_S)$ comes from the interference between the bottom two Feynman diagrams and the subset of the first diagram in the upper row. It is important to note that the gluon-initiated diagrams do not interfere with the bottom two diagrams.

\begin{table}[tb!]
\begin{center}
\resizebox{1.0\columnwidth}{!}{%
 \begin{tabular}{|c|c||p{0.2\textwidth}|p{0.2\textwidth}|c|}
\hline
\multirow{2}{1em}{} &  $\sigma(p p \rightarrow \Psi \bar{\Psi})$ (fb) & \multicolumn{3}{c|}{$\sigma(p p \rightarrow \Psi \bar{\Psi})$ (fb) for} \\
\cline{2-2}
\multirow{2}{1em}{BP}&  LO  & \multicolumn{3}{c|}{leading production processes at LO and NLO} \\
\cline{3-5}
&  $\sigma_{\text{LO}}= \mathcal{O}(\alpha_S^2)+\mathcal{O}(f^4)+\mathcal{O}(f^2\alpha_S)$ & LO, $~~\mathcal{O}(\alpha_S^2)$   & NLO, $~~\mathcal{O}(\alpha_S^3)$   &K-fac\\
\hline\hline
BP1  & $96.39^{+31.5\%}_{-22.5\%} $ & $105.8^{+31.3\%}_{-22.2\%}  $ & $138.5^{+9.6\%}_{-11.3\%} $ & 1.31 \\
\hline\hline
BP2  & $115.2^{+32.7\%}_{-23.2\%} $ & $125.7^{+31.4\%}_{-22.4\%}  $ & $162.1^{+10.1\%}_{-11.5\%} $ & 1.29 \\
\hline\hline
BP3  & $185.6^{+32.4\%}_{-23.0\%} $ & $201.6^{+31.3\%}_{-22.3\%}  $ & $257.3^{+9.8\%}_{-11.4\%} $ & 1.28 \\
\hline\hline
 \end{tabular} 
 }
\caption{Total leading-order cross section, including QCD and BSM coupling, and their interference in the pair production of VLQ at the 14 TeV LHC before their decay is given in the left panel. Right panel: Leading contribution of the tree-level VLQ pair production process ($\mathcal{O}(\alpha_S^2)$) and its next-to-leading order cross section, along with the integrated K-factor, are given. The superscript and subscript denote the scale uncertainties (in percentage) of the total cross section. Five massless quark flavors are used for computation.}
\label{tab:crosssection_1_5}
\end{center}
\end{table}

\begin{figure}[tb!]
\centering
\includegraphics[scale=0.35]{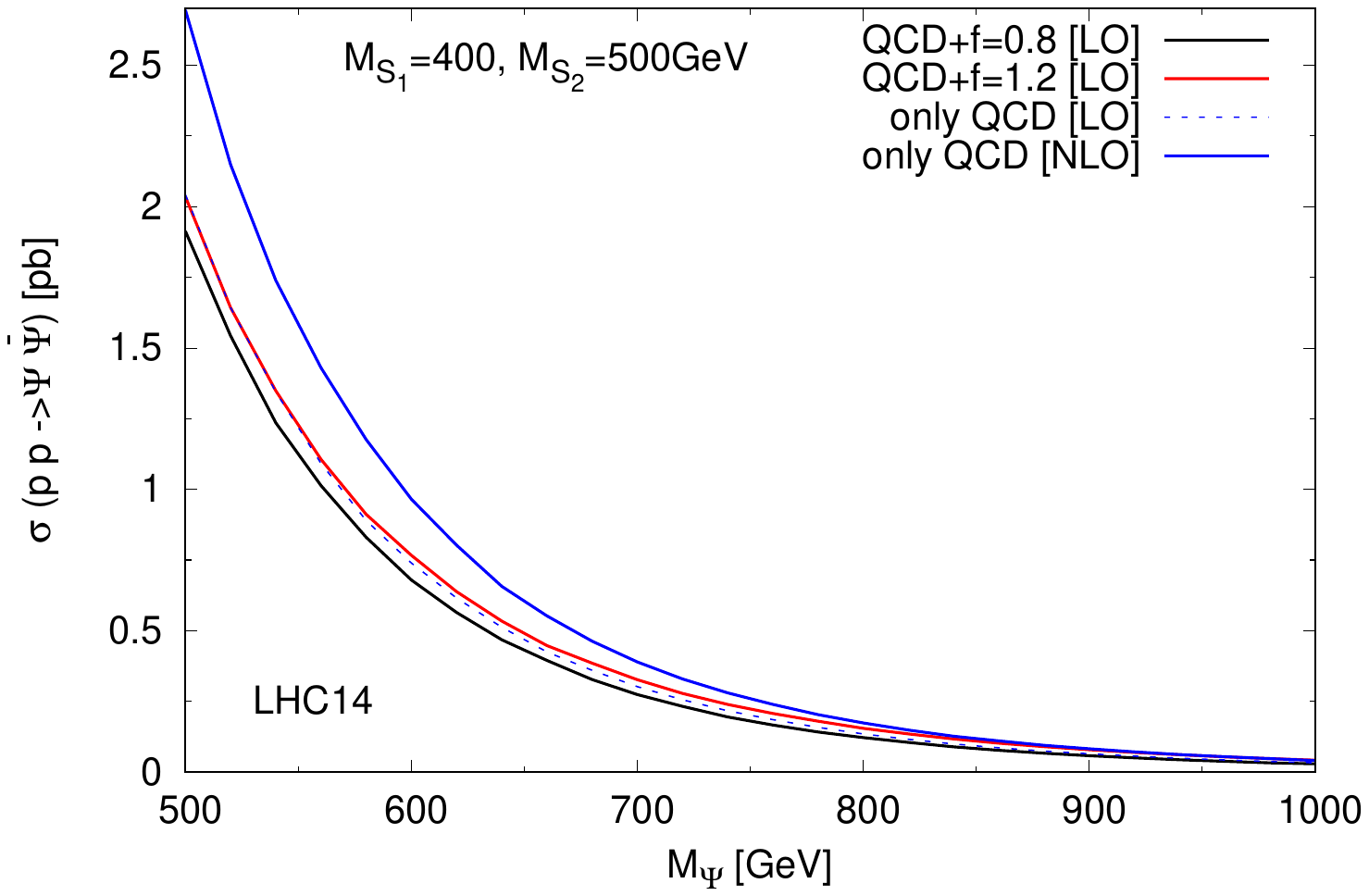}
\caption{This plot illustrates the variation of the vector-like quark ($\Psi$) pair production cross-section with VLQ mass at the 14 TeV LHC, focusing on its coupling with the first quark family ($f_u\neq 0, f_c=0$). The red (black) line depicts the total leading-order (LO) cross-section, including pure QCD, Yukawa interactions, and their interferences, for $f_u = 1.2~ (0.8)$. The blue dashed line shows the leading-order contributions solely from pure QCD processes for $\Psi \bar{\Psi}$ production, while the solid blue line represents the next-to-leading-order (NLO) QCD cross-section for pure QCD processes. Scalar masses are fixed at $M_{S_1} = 400$ and $M_{S_2} = 500$ GeV. }
\label{rates}
\end{figure}
%

\begin{figure}[tb!]
\centering
  \subfloat[] {\label{fig_logPt_5}\includegraphics[width=0.495\textwidth]{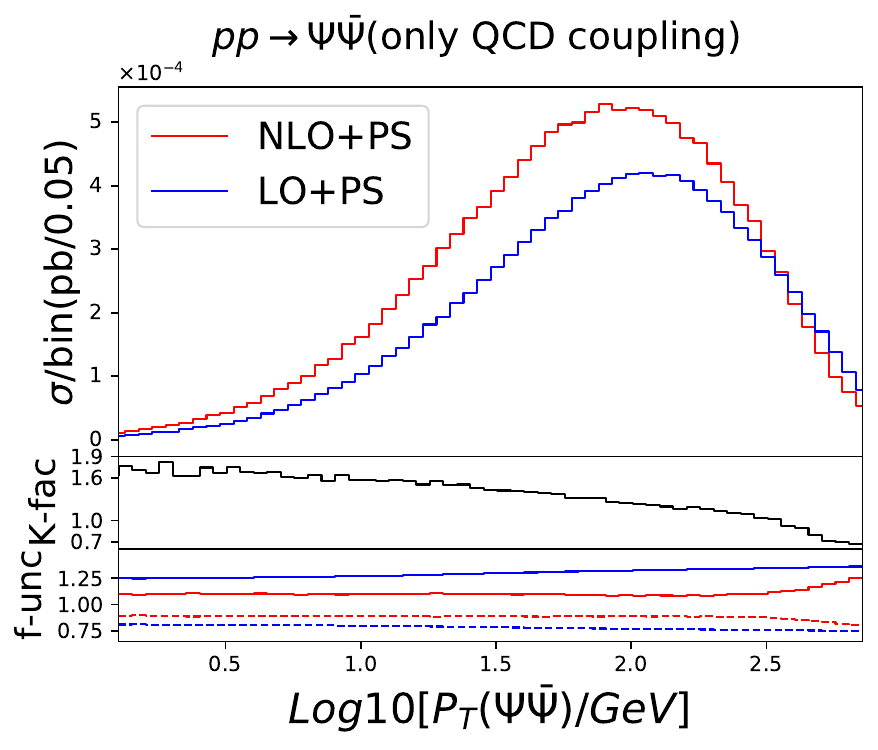}} 
  \subfloat[] {\label{fig_invMass_5}\includegraphics[width=0.495\textwidth]{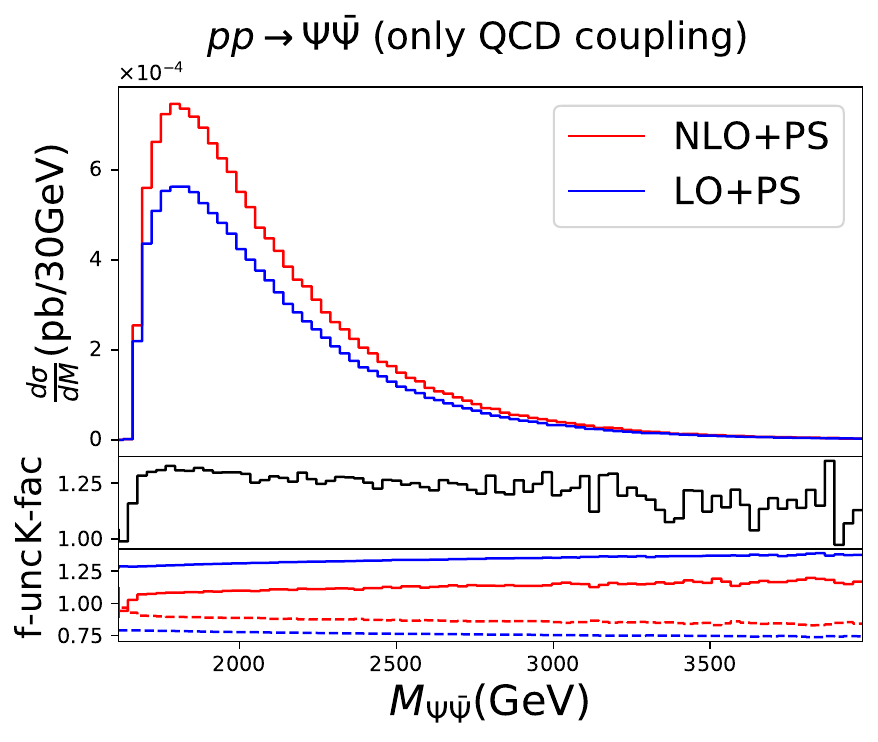}}
\caption{(a) Distribution of $\log_{10}[P_T(\Psi\bar{\Psi})/GeV]$ at LO+PS and NLO+PS for VLQ pair production at the 14 TeV LHC (upper panel). (b) The distribution of the invariant mass of the VLQ pair (upper panel). The differential K-factor and the scale uncertainties are shown in the middle and bottom panels, respectively. The plots correspond to BP1, and LO consists only of QCD coupling.}
\label{fig_LOvsNLO_5}
\end{figure}


The leading production channels (three diagrams at the top row) form a gauge invariant subset, and we do one-loop QCD correction of those processes. As a result, the NLO cross section has the order $\mathcal{O}(\alpha_S^3)$. Few representative Feynman diagrams at NLO-QCD are shown in Figure \ref{FD_production_NLO-1_5}. The total LO cross section is given in the left panel of Table \ref{tab:crosssection_1_5}. The leading contribution of the VLQ pair production at the tree level and its next-to-leading order cross section, along with the integrated K-factor, are given in the right panel of Table \ref{tab:crosssection_1_5}. The K-factor is defined as the ratio of NLO to LO cross section. We find a significant enhancement of about $30\% $ in the NLO-QCD cross-section over LO. \\

Fig.~\ref{rates} illustrates the production cross-section $\sigma(pp\rightarrow \Psi \bar{\Psi})$ at the 14 TeV LHC for coupling with the first families ($f_u\neq 0, f_c= 0$). The total LO cross-section is depicted by the solid red (black) line, originating from pure QCD, t-channel Yukawa diagrams, and their interferences with Yukawa coupling $f_u = 1.2~ (0.8)$. The blue dashed line represents the LO cross-section from pure QCD diagrams, while its NLO-QCD correction is depicted by a solid blue line. The contributions from the QCD-induced diagrams do not depend on the dark matter mass and the Yukawa coupling. In the case of considering second families ($f_u =0, f_c\neq 0$), the contribution from t-channel Yukawa diagrams and their interference with the QCD diagram is exceedingly tiny due to the parton density function. Consequently, the total LO cross-section is nearly identical to the pure QCD diagrams. Interestingly, in Fig.~\ref{rates}, there is a destructive interference between the QCD and t-channel Yukawa diagrams. Excluding constant numerical factors, the squared matrix elements for $u\bar{u}$-initiated processes at LO are as follows. 
\begin{equation}
\begin{split}
& |\mathcal{M}_{QCD}^{u\bar{u}}|^2\propto \alpha_s^2 C_A C_F \dfrac{2 M_\Psi^4 +t^2+2 M_\Psi^2(s-t-u)+u^2}{s^2}~, \\ &
|\mathcal{M}_{S_{1,2}}|^2\propto f^4 \dfrac{(M_\Psi^2-t)^2}{(M_{S_{1,2}}^2-t)^2}~.
\end{split}
\label{square_matrix}
\end{equation}
The interference terms are as follows.
\begin{equation}
\begin{split}
& 2 ~Re( \mathcal{M}_{QCD}^{u\bar{u}} \mathcal{M}_{S_{1,2}}^*) \propto - \alpha_s f^2 C_A C_F \dfrac{M_\Psi^4 + M_\Psi^2(s-2t)+t^2}{2s (M_{S_{1,2}}^2-t)}~, \\ &
2 ~Re( \mathcal{M}_{S_1} \mathcal{M}_{S_2}^*) \propto f^4 \dfrac{(t-M_\Psi^2)^2}{(t-M_{S_2}^2) (t-M_{S_1}^2)}~.
\end{split}
\label{square_matrix_inter}
\end{equation}
$\mathcal{M}_{QCD}^{u\bar{u}}$ represents the amplitude of the top left diagram in Fig.~\ref{FD_production_5} with the initial state $u\bar{u}$, while $\mathcal{M}_{S_1}$ and $\mathcal{M}_{S_2}$ denote the amplitudes of the bottom two diagrams in Fig.~\ref{FD_production_5}. The Mandelstam variables are represented by $s,~t,~u$, and constants $C_A=3$ and $C_F=4/3$. 
According to Eq.~\ref{square_matrix_inter}, destructive interference can occur between the Yukawa and pure QCD diagrams. At moderate Yukawa couplings, with $f\leq 0.8$, the destructive interference reduces the total cross-section. The production rate exceeds the contribution from pure QCD only in larger Yukawa coupling regimes, specifically when $f\geq 1.2$. Due to the dominance of QCD contributions, we analyze pure QCD-initiated diagrams at LO and their NLO-QCD corrections in the subsequent collider analysis.

We designate the partonic centre-of-mass energy of the event as the central choice for both the factorization and renormalization scales. To compute the scale uncertainties, we vary the factorization and renormalization scales from two to half of this central scale, resulting in nine different data sets. The superscripts and subscripts in the tables indicate the envelopes of the nine data sets, although all of the cross sections shown in Table \ref{tab:crosssection_1_5} correspond to the central scale. At NLO, the uncertainties associated with the renormalization and factorization scales are significantly smaller than at LO.

LO+PS and NLO+PS distributions of  $\log_{10}[P_T(\Psi\bar{\Psi})/GeV]$ (upper panel) and the differential K-factor (middle panel) are given in Figure~\ref{fig_logPt_5}. $P_T(\Psi\bar{\Psi})$ is the transverse momentum of the VLQ pair. For $\log_{10}[P_T(\Psi\bar{\Psi})/GeV] < 1.6$, the left plot shows that K-factor is more than 1, but for $\log_{10}[P_T(\Psi\bar{\Psi})/GeV] > 1.6$, the K-factor is less than 1, indicating that the NLO cross section is less than the LO cross section. The differential K-factor is not flat everywhere. It is almost flat at the lower values and then starts to go down, so scaling the LO events by a constant K-factor would not give accurate results.

The invariant mass distribution of the VLQ pair is shown on the top panels of Figure \ref{fig_invMass_5} for BP1. The differential K-factor is shown in the middle panel. Invariant mass distribution peaks around 1800 GeV, and the differential K-factor is almost flat around the peak. The bottom panels show the envelope of the factorization and renormalization scale uncertainties. The red solid and dashed lines show the width of the scale uncertainty for NLO+PS, while the blue solid and dashed lines show the LO+PS scale uncertainties. We can see that both LO+PS and NLO+PS results are stable, but the NLO+PS result has much-reduced scale uncertainty. Although these are figurative findings because the decay of $\Psi$ is not considered, they demonstrate the need to do $\mathcal{O}(\alpha_s)$ corrections on the pair production channels for more accurate prediction of the total cross section, differential distribution of various variables, and lower scale uncertainty. 

\section{Multivariate Analysis (MVA)}
\label{collider_5}


We conducted a collider analysis on this model, focusing on the $t\bar{t}+\text{MET}$ final state\footnote{Multivariate analysis of fatjets plus significant missing energy as the final state employing jet substructure variables are frequently searched in the context of the LHC for many BSM scenarios~\cite{Ghosh:2023ocz,  PhysRevD.105.115038}.}. To generate the signal, we utilized NLO+PS accurate events. The branching ratio of the VLQ decays into a top quark associated with the scalar is less than 0.5, as shown in Table \ref{parameters_5}. Representative benchmark points listed in Table \ref{parameters_5} are consistent with the 139 $\text{fb}^{-1}$ projected exclusion contour from a recent ATLAS analysis \cite{ATLAS:2017eoo} of stop pair production. The signal topology is given by 
\footnote{Since we are considering two top-like fatjet ($J_t$) without measuring the jet charge, $uu\rightarrow\Psi \Psi$ (through the t-channel scalar mediator) can also contribute to the same signature followed by the decay of the VLQ into the top. Interestingly, since scalars and $\Psi$ have the same PQ charge, this type of t-channels exchange is impossible unless PQ symmetry is spontaneously broken finally to contribute negligibly. 
 To give some perspective in our benchmark point BP1, we find $\sigma_{\text{LO}}(u u\rightarrow \Psi\Psi)= 0.3~\text{fb}$, and $\sigma_{\text{LO}}(\bar{u} \bar{u}\rightarrow\bar{\Psi}\bar{\Psi})= 0$, so we safely ignore those processes.},
\begin{equation}
p p \rightarrow \Psi \bar{\Psi} \hspace{1mm} \text{[QCD]} \rightarrow (t \hspace{1mm} S_{1,2}) (\bar{t} \hspace{1mm} S_{1,2}) \hspace{1mm} j \Rightarrow 2 J_t +\slashed{E}_T + X~.
\label{Eq-topology_5}
\end{equation}
The advantages of studying the hadronic final state are as follows. The large hadronic branching ratio and significant mass difference between VLQ and the scalar significantly boost the top quark. As a result, reconstructing the top quark as a boosted fatjet additionally provides the inherent properties of the jet. Furthermore, we can have additional handel using jet substructure variables.

\begin{table}[tb!]
\begin{center}
 \scriptsize
 \setlength\tabcolsep{2.7pt} 
\begin{tabular}{@{}*{11}{|p{.084\textwidth}@{}}|}
\hline
& \textbf{Signal (BP1)} & \textbf{$Z$+jets}  & \textbf{$W$+jets} & \textbf{$t \bar{t}$+jets} & \textbf{$tW$+jets} & \textbf{$WZ$+j}& \textbf{$WW$+j}  & \textbf{$ZZ$+j} & \textbf{$t\bar{t}~V$} & \textbf{tot BG}  \\
& (fb)  & (fb) & (fb) & (fb) & (fb) & (fb) & (fb) & (fb) & (fb) & (fb) \\
\hline \hline
\multirow{2}{*}{C1} & 5.99  & 2517.99  & 1366.91 & 690.65
& 366.91 & 93.53 & 25.90 & 11.51 & 8.34 & 5081.74\\
& [$100\%$]  & [$100\%$] & [$100\%$] & [$100\%$] 
& [$100\%$] & [$100\%$] & [$100\%$] & [$100\%$] & [$100\%$] & [$100\%$] \\
\hline
\multirow{2}{*}{C2} & 5.49  & 1640.29  & 762.59 & 302.16
& 152.52 & 58.35 & 11.51 & 6.973 & 6.17 & 2940.56\\
& [$91.65\%$]  & [$65.14\%$] & [$55.79\%$] & [$43.75\%$] 
& [$41.57\%$] & [$62.39\%$] & [$44.44\%$] & [$60.58\%$] & [$73.98\%$] &[$57.87\%$] \\
\hline
\multirow{2}{*}{C3}  & 4.58 & 241.73  & 117.99 & 230.94
& 114.39 & 10.79 & 2.45 & 1.92 & 5.11 & 725.32\\
& [$76.46\%$]  & [$9.60\%$] & [$8.63\%$] & [$33.44\%$] 
& [$31.18\%$] & [$11.54\%$] & [$9.46\%$] & [$16.69\%$] & [$61.27\%$] &[$14.27\%$] \\
\hline
\multirow{2}{*}{C4}  & 2.23 & 25.38  & 17.33 & 64.23 
& 27.45 & 1.24 & 0.33 & 0.2 & 2.30 & 138.46\\
& [$37.23\%$] & [$1.01\%$] & [$1.27\%$] & [$9.30\%$] 
& [$7.48\%$] & [$1.33\%$] & [$1.27\%$] & [$1.74\%$] & [$28.13\%$] &[$2.72\%$] \\
\hline
 \end{tabular}     
\caption{After applying various kinematic event selection cuts, signal and background events (in fb) indicate the efficiency for each set of cuts to reduce the backgrounds. The kinematic cuts (C1-C4) are described in the text. After applying the C4 cut, the remaining events are passed for the multivariate analysis. }
\label{tab:cut-flow_5}
\end{center}
\end{table}

\begin{table}[tb!]
\centering
\resizebox{\columnwidth}{!}{
 \begin{tabular}[b]{|c|c|c|c|c|c|c|c|c|c|c|c|c|}
\hline
& $\slashed{E}_T$ & $\Delta \phi(J_1,\slashed{E}_T)$  & $\Delta R (J_0, J_1)$ & $\tau_{32}(J_1)$ &  $\tau_{32}(J_0)$   & $\Delta \phi(J_0,\slashed{E}_T)$ & $\text{M}_{\text{eff}}$ & $M(J_1)$  & $M(J_0)$ & $\tau_{31}(J_0)$ & $\tau_{31}(J_1)$ \\ %
\hline\hline
BP1 &31.29 & 20.19 & 17.82 & 8.61 & 8.49 & 8.38 & 3.29 & 2.10 & 1.48 & 1.10 & 0.9 \\
\hline
BP2 & 19.39 & 16.74 & 17.39 & 6.75 & 6.99 & 8.04 & 2.26 & 0.67 & 0.72 & 1.11 & 0.73 \\
\hline
BP3 & 9.25 & 11.30 & 12.11 & 6.52 & 5.74 & 6.55 & 1.29 & 0.95 & 0.38 & 0.52 & 0.72 \\
\hline
 \end{tabular} 
 }
\caption{Method unspecific relative separation power of different kinematic variables in separating the signal and background classes.}
\label{relative_imp_5}
\end{table}
%
\textbf{Simulation details:} Parton-level events are generated using {\sc MadGraph5\_aMC@NLO} \cite{Alwall:2014hca} and subsequently processed through {\sc Pythia8} \cite{Sjostrand:2001yu, Sjostrand:2014zea} for showering, fragmentation, and hadronization. Background events are generated, incorporating two to four additional jets with {\sc MLM} matching \cite{Mangano:2006rw, Hoeche:2005vzu} and virtually ordered Pythia showers to prevent double counting. The showered events are then subjected to {\sc Delphes3} \cite{deFavereau:2013fsa} to account for detector effects using the default CMS card. We construct $\mbox{anti-k}_T$ \cite{Cacciari:2008gp} jets utilizing particle-flow towers and particle-flow tracks as input. Large radius fatjets of radius 1.5 are formed using the Cambridge-Aachen (CA) algorithm \cite{Dokshitzer:1997in} with {\sc Fastjet 3.2.2} \cite{Cacciari:2011ma}. Finally, the multivariate analysis (MVA) is conducted using the adaptive Boosted Decision Tree (BDT) algorithm within the {\sc TMVA} \cite{Hocker:2007ht} framework.\\

\textbf{Standard Model backgrounds:} Our analysis accounts for all relevant backgrounds significantly contributing to the two boosted top fatjets with substantial missing transverse momentum, outlined below.\\

\noindent
\underline{$t \bar{t}+$ jets:} The dominant background in our signal process is $t \bar{t}+$ jets, primarily from semi-leptonic top decays. Pure hadronic decay of tops is reduced significantly due to the requirement of substantial missing energy. In semi-leptonic decay, one top is reconstructed as a boosted top jet, while the other contributes missing energy when the lepton goes undetected. Additional boosted jets come from QCD radiation. This background is matched using the MLM matching scheme with up to two extra jets.\\

\noindent
\underline{$t W+$ jets:} Single top quark production associated with the W boson contributes significantly to the SM background. The top is reconstructed as a boosted fatjet, and the W boson decays leptonically, leading to missing transverse momentum. Another boosted fatjet arises from QCD jets, with MLM matching up to two extra jets applied.\\

\noindent
\underline{$V+$ jets:} At the LHC, each of the $Z+$ jets and $W+$ jets processes has a cross-section of $\sim 10^4~pb$ \cite{Catani:2009sm, Balossini:2009sa}. Leptonic decay of $W$ (with the lepton going undetected) and the invisible decay of the $Z$ boson result in a large MET, while the initial state QCD radiations mimic the fatjets. Due to the substantial cross-section, $V+$ jets ($V=Z, W$) can contribute significantly even with requirements for a large ($\sim m_{\text{top}}$) reconstructed mass of the fatjets and b-tagging within either of the fatjets. MLM matching is applied for both processes, considering up to four extra jets.\\

\noindent
\underline{$V V+$ jets}. Minor contribution can come from di-boson $+$ jets. We retain all the three di-boson background processes ($pp\rightarrow W Z, W W,Z Z$) in our analysis. Among the three, $W Z+$ jets contribute the most. All three processes are matched up to two extra jets with an MLM matching scheme. In all the cases, one of the vector boson decay invisibly ($Z\rightarrow \nu \nu$) or leptonically ($W \rightarrow l \nu $), contributing to MET. One boosted fatjet originates from QCD jets, while another fatjet arises from either the hadronically decaying vector boson or QCD jets.\\

\noindent
\underline{$t \bar{t} V$ :} Processes like $t \bar{t} V$ ($V=Z, W$) have three-body phase spaces and lower cross-sections compared to the background processes mentioned above, but they exhibit signal-like characteristics. Hence, we retain this background in our analysis. Both tops can be reconstructed as boosted fatjets, with MET originating from the invisible or leptonically decaying $Z$ and $W$ bosons.\\
In our analysis, we normalize all background processes using the available $NLO/NNLO$ QCD cross-section.\\ 

The expected number of signal (BP1) and background events (in fb, expected event numbers are obtained by multiplying them with the luminosity) is listed as cut flow, along with the cut efficiencies, after each set of event selection criteria is shown in Table \ref{tab:cut-flow_5}. In preselection cut (C1) we demand at least two fatjets of radius $R=1.5$, each with a transverse momentum $P_T(J_0), P_T(J_1)> 200$ GeV, missing transverse momentum $\slashed{E}_T>100$ GeV, a lepton-veto, and $|\Delta \Phi(J_{0,1},\slashed{E}_T)|>0.2$ (to minimize jet mismeasurement contribution to $\slashed{E}_T$). The other cuts are: 
(C2) $\slashed{E}_T>150$ GeV, 
(C3) a b-tag within the leading or subleading fatjet, and 
(C4) pruned mass of the two leading jets $ M_{J_0}, M_{J_1}>120$ GeV. After applying the preselection cut (C1), we find $V$+jets ($V = Z, W$) are the principal background while $t \bar{t}$+jets is the subdominant background. However, after a b-tag within $J_0$ or $J_1$ and demanding large fatjet masses, we found $t \bar{t}$+jets becomes the primary background, while $V$+jets are the subdominant. Applying all those cuts, we still retain a substantial number of signal events while the background reduces significantly.
%
\begin{figure}[tb!]
\centering
  \subfloat {\label{correlation_sig_5}\includegraphics[width=0.495\textwidth]{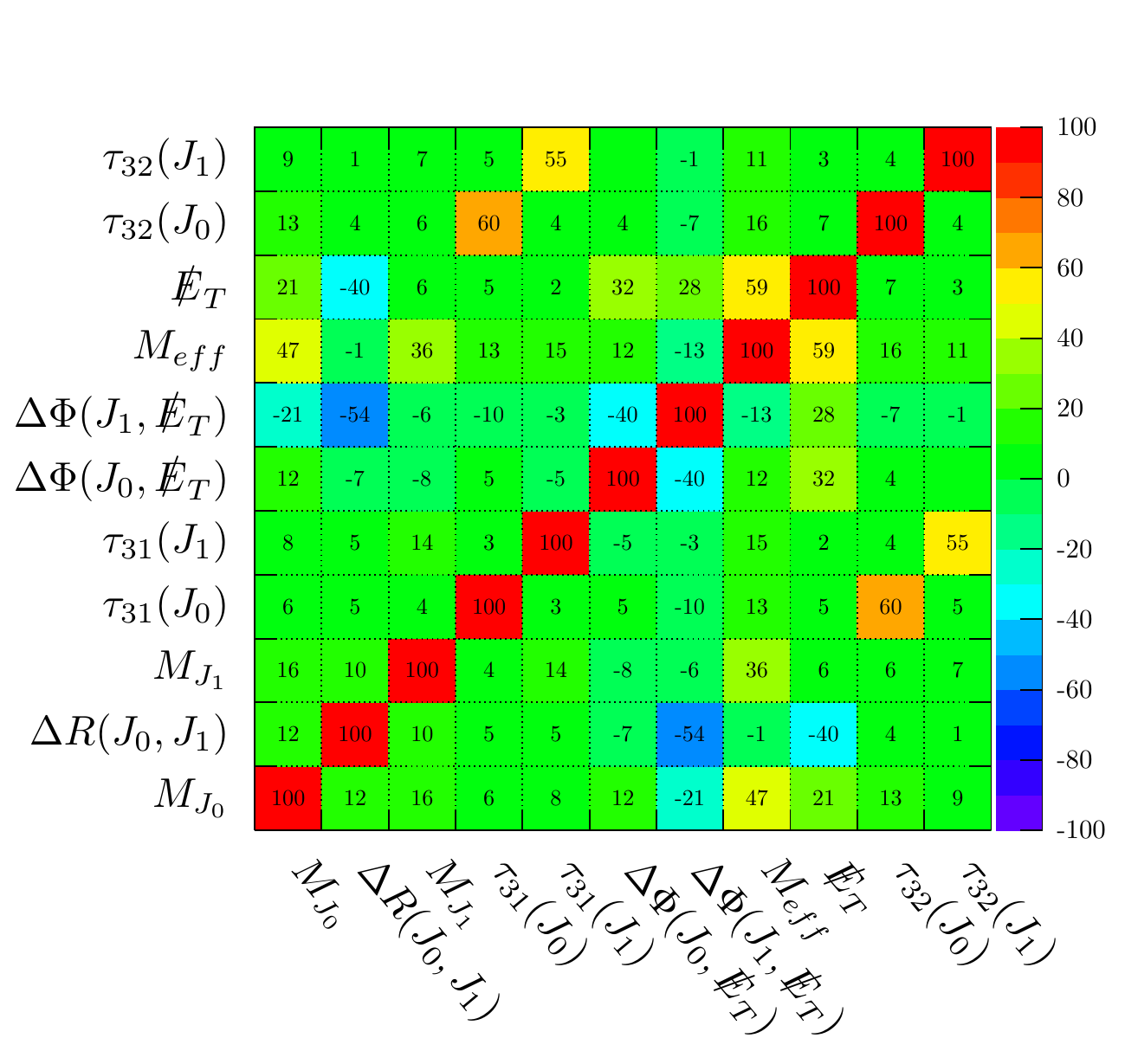}} 
  \subfloat {\label{correlation_bg_5}\includegraphics[width=0.495\textwidth]{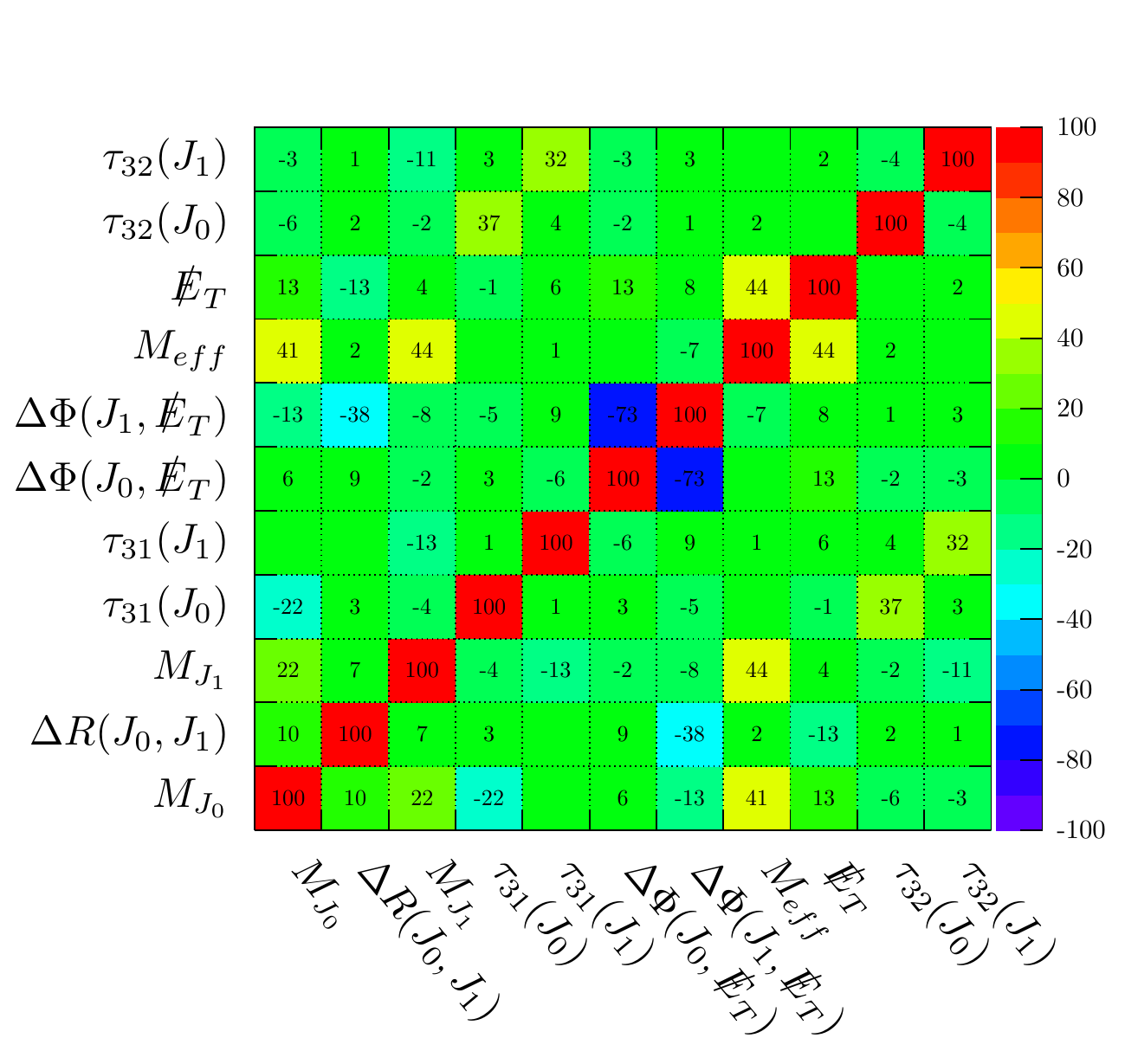}}
  \caption{Coefficients of linear correlation (in percentage) between various kinematical variables for the signal (BP1, left panel) and background (right panel) are presented. Missing entries have an insignificant correlation of less than one. Two variables are correlated or anti-correlated based on positive and negative coefficients.}
  \label{Fig:correlation_5}
\end{figure}
\begin{table}[tb!]
\begin{center}
 \begin{tabular}{|c c|}
\hline
BDT hyperparameters & Optimal selection \\ 
\hline\hline
BoostType & AdaBoost  \\
AdaBoostBeta & 0.5  \\
NTrees & 165  \\
MinNodeSize & $5.6\%$  \\
MaxDepth & 3  \\
UseBaggedBoost & True  \\
BaggedSampleFraction & 0.5  \\
SeparationType & GiniIndex  \\
nCuts & 20  \\

\hline
 \end{tabular} 
\caption{ Optimized BDT hyperparameters}
\label{tab:opti_param}
\end{center}
\end{table}
%
All the signal and background processes are passed through all these event selection criteria up to C4 before passing events to MVA. We create two separate signal and background classes. The combined background is the weighted combination of all the different background processes. Each signal and background class is randomly divided into $50\%$ for training and the rest $50\%$ for testing. We use boosted decision tree (BDT) algorithm and choose a set of kinematic variables from a wider collection of variables for MVA. The variables with high relative importance distinguishing the signal class from the background class are preferable. Table \ref{relative_imp_5} lists the relative importance of the various kinematic variables involved in the MVA. The left (signal) and right (background) tables of Figure \ref{Fig:correlation_5} show the linear correlation coefficients among the variables employed in MVA for BP1. The optimized BDT hyperparameters used in our analysis are outlined in Table \ref{tab:opti_param}.

\begin{figure}[tb!]
\centering
  \subfloat {\label{output_5}\includegraphics[width=0.495\textwidth]{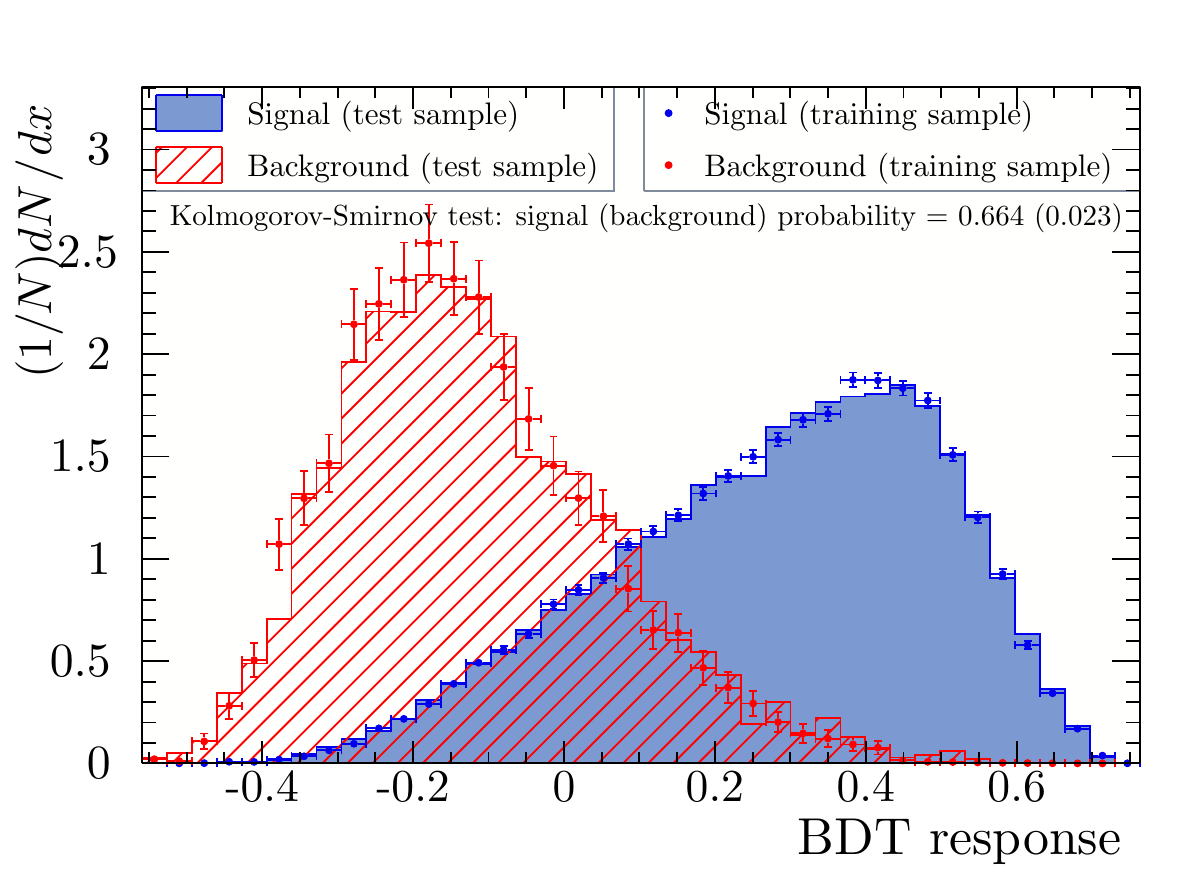}} 
  \subfloat {\label{significance_5}\includegraphics[width=0.495\textwidth]{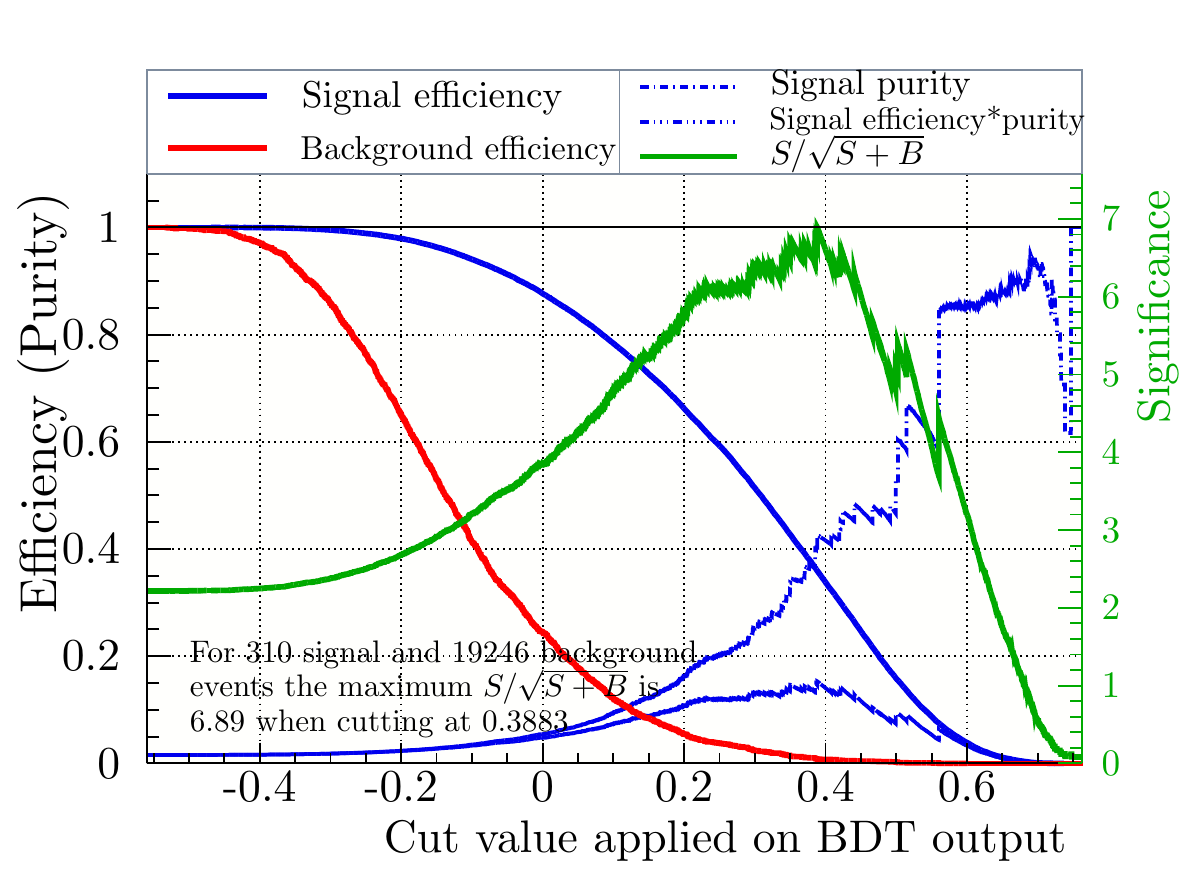}}
  \caption{The left panel shows the normalized distribution of the BDT output for training and testing samples of both signal and background. The statistical significance of the signal with the cut applied to the BDT output is shown in the right panel, along with signal and background efficiency.}
\label{plot:MVA-Output-Signi_5}
\end{figure}

\begin{table}[tb!]
\begin{center}
 \scriptsize
 \begin{tabular}{|c|c|c|c|c|c|c|c|}
\hline
 & $N_S^{bc}$ (fb) & BD$T_{opt}$ & $N_S$ (fb) & $N_B$ (fb) & $\frac{N_S}{\sqrt{N_S+N_B}}$ &  $\frac{N_S}{\sqrt{N_S+N_B+(0.05\times N_B)^2}}$  & $\frac{N_S}{N_B}$  \\ 
  &  &  &  &  &  139 (300) $\text{fb}^{-1}$  &  ( 300 $\text{fb}^{-1}$ ) & \\ 
\hline\hline
BP1 & 2.23 & 0.3883 & 0.8012 & 1.0783 & 6.89 (10.12) & 8.37 & 0.743 \\
\hline
BP2 & 2.53 & 0.2582 & 1.2207 & 6.1302 & 5.31 (7.80) & 3.55  & 0.199 \\
\hline
BP3 & 1.67 & 0.2961 & 0.4252 & 2.3529 & 3.0 (4.42) & 2.80  & 0.180 \\
\hline
$N_{SM}$ & 138.46 & \multicolumn{6}{| c |}{} \\
\hline
 \end{tabular} 
\caption{The table shows the efficacy of the current search in terms of statistical significance for various benchmarks. $N_S^{bc}$ and $N_{SM}$ are the total numbers of signal and background events before performing MVA (see Table \ref{tab:cut-flow_5}), while $N_S$ and $N_B$, respectively, provide those following BDT analysis. BD$T_{opt}$ is the optimal BDT cut. The sixth column shows the signal's statistical significance for 139 (300) $\text{fb}^{-1}$ luminosity, including Delphes simulation without systematics. The following column presents an approximate statistical significance with a $5\%$ systematic consideration. The last column illustrates the signal-to-background ratio.
}
\label{tab:MVA-result_5}
\end{center}
\end{table}
 
The normalized distribution of the BDT response for test and train samples of both signal (BP1) and background classes is plotted on the left side of Figure \ref{plot:MVA-Output-Signi_5}. We find signal and background are well separated. With the cut applied to the BDT output, the signal and background efficiency, as well as the statistical significance ($\frac{N_S}{\sqrt{N_S+N_B}}$) for 139 $\text{fb}^{-1}$ data, are presented in the right plot of Figure \ref{plot:MVA-Output-Signi_5}. Before applying any cuts to the BDT output, Table \ref{tab:MVA-result_5} shows the number of signals ($N_S^{bc}$) and background ($N_{SM}$) events for various BPs. It also shows the expected number of signal events ($N_S$) and background events ($N_B$) that remain after applying an optimal cut (BD$T_{opt}$) to the BDT output. We optimize each of the three BPs separately. The sixth column displays the statistical significance of the signal at 139 (300) $\text{fb}^{-1}$ luminosity, incorporating Delphes simulation without systematics. The subsequent column provides an estimated statistical significance, accounting for a $5\%$ systematic consideration. The last column illustrates the signal-to-background ratio. 

Table \ref{tab:MVA-result_5} shows that the statistical significance of BP3 is lower than that of the other two benchmark points even though it has the most significant partonic cross section of VLQ pair production since the mass of the VLQ is the smallest for BP3. This is attributed to a smaller mass difference between VLQ and DM than the other two BPs, which results in a relatively less boosted top quark and a smaller signal efficiency. Table \ref{tab:MVA-result_5} also demonstrate that a significant parameter space of this model can be explored with more than $5\sigma$ significance using the 300 $\text{fb}^{-1}$ data at the 14 TeV LHC. The exclusion reach ($2\sigma$) of this model at the HL-LHC ($3000~\text{fb}^{-1}$) extends up to VLQ masses of 1253 GeV, 1188 GeV, and 1038 GeV for mass gaps between the VLQ and dark matter of 500 GeV, 407 GeV, and 300 GeV, respectively, assuming a branching ratio of 0.5 for VLQ decays into the top quark associated with the scalar. Consequently, the HL-LHC can exclude a significant portion of the allowed parameter space in Figure \ref{scan_plot_5}.

\section{Conclusions}
\label{sec:conclsn_5}

We explore a complex scalar extended KSVZ axion framework, where the scalar is singlet under the SM gauge groups but only has the Peccei-Quinn charge. This model has the capability to solve two of the most outstanding problems of the SM, that is, the strong-CP problem and a natural candidate for dark matter in the form of QCD axion having a lifetime comparable to the age of the Universe. Axion can satisfy the correct dark matter relic density, measured by the Planck collaboration, but at the expense of fine-tuning the corresponding breaking scale. The residual $\mathbb{Z}_2$  symmetry in this model ensures that the lightest component of the complex scalar is stable and thus plays the role of a second dark matter, removing the need for any such fine-tuning.

KSVZ axion framework also provides a rich phenomenology by introducing a vector-like quark which can be explored at a hadron collider like the LHC. In the extended scenario, VLQ interacts with the scalar (DM) candidate and the SM quarks (up or down) based on its hypercharge. Hence the VLQ now plays a critical role in dark matter phenomenology because it opens up new annihilation and coannihilation channels.

Here, we explore the possibility of democratic Yukawa interaction of the vector-like quark with all up-type quarks and scalar dark matter candidate. One must find the allowed parameter spaces that provide the correct relic density and agree with other experimental observations such as direct detection (DD), collider data, etc. It is found that the flavor constraint strongly disfavours this democratic option, which requires either one or both lighter flavor couplings ($f_u, f_c$) needs to be tiny. For simplicity, we consider $f_c = 0$ while keeping the other two democratic. 
It is interesting to note that the allowed parameter space can neither support arbitrarily large coupling $f (= f_u = f_t)$ from direct detection nor the too-small value of it to obtain the correct relic density. Therefore, their interplay remains vital for selecting the available parameter spaces.

We employ NLO-QCD correction on dominant production channels of colored VLQ pair production at the LHC. The total NLO cross section increases by approximately $30\%$ compared to LO. Additionally, the differential distributions of various observables exhibit significant changes when considering NLO+PS compared to LO+PS. A notable reduction in scale uncertainty is also observed at  NLO+PS, leading to a more precise and accurate result than at LO+PS. Following pair production at the LHC, each VLQ undergoes decay into a top quark accompanied by a scalar DM. The VLQ and scalar DM exhibit a substantial mass difference, considerably boosting both top quarks. Boosted top-like fatjets generated from the hadronic decay of top quark still carry different characteristics of it, which are primarily captured in a dedicated jet analysis and different substructure variables. Multivariate analysis with these variables and attributes of event topology is demonstrated to establish a strong ability to explore a significant parameter space of this model at the 14 TeV LHC.

\acknowledgments
We thank Dr. Satyajit Seth for the fruitful discussions. This work is supported by the Physical Research Laboratory (PRL), Department of Space, Government of India. The Computations were performed using the Param Vikram-1000 HPC and TDP project at PRL.

\appendix
\section{Appendix}
\label{appen_5}

\begin{figure}[htbp!]
\centering
\includegraphics[width=0.32\textwidth]{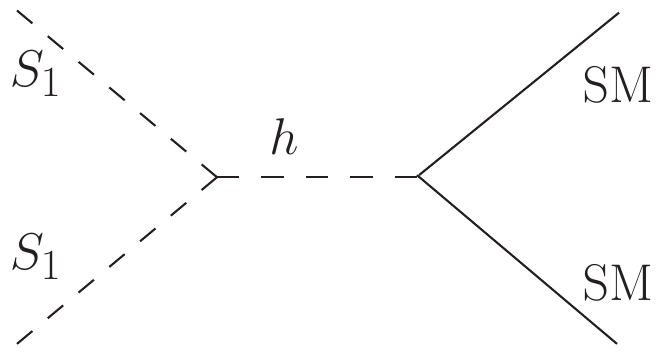}
\includegraphics[width=0.32\textwidth]{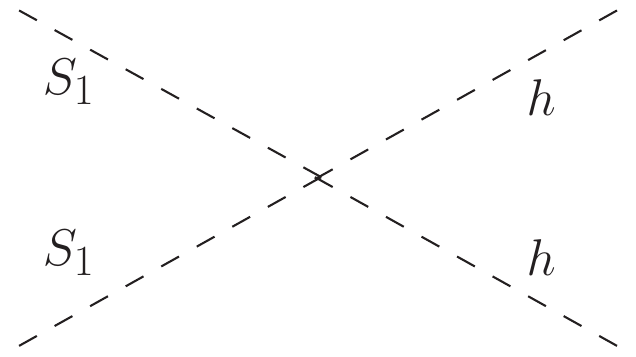}
\includegraphics[width=0.32\textwidth]{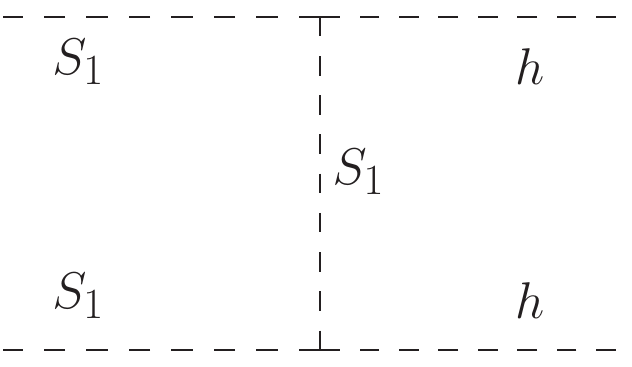}\\
\includegraphics[width=0.32\textwidth]{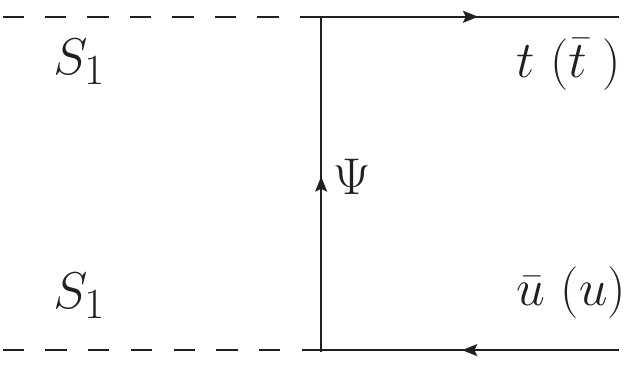}\qquad
\includegraphics[width=0.32\textwidth]{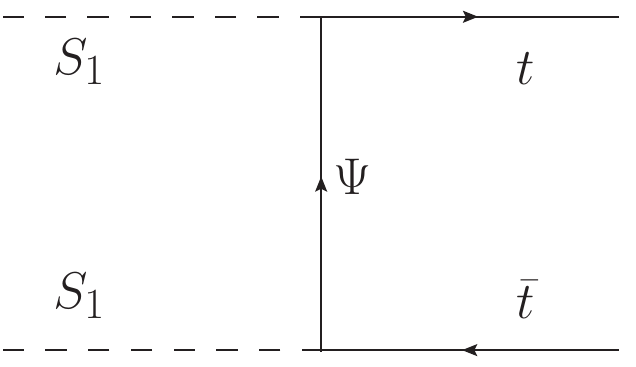}
\caption{Feynman diagrams of scalar dark matter annihilation into Standard Model particles are presented. }
\label{annihilation_5}
\end{figure}
\paragraph{Direct detection channels:}
\begin{figure}[htbp!]
\centering
\subfloat[$\mathcal{M}_1$] {\label{DD_s} \includegraphics[width=0.32\textwidth]{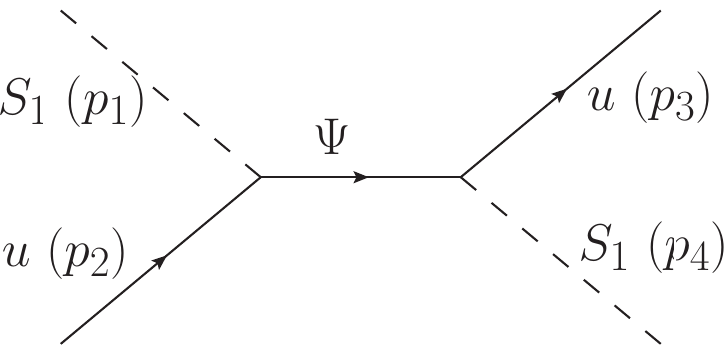}}
\subfloat[$\mathcal{M}_2$] {\label{DD_t} \includegraphics[width=0.32\textwidth]{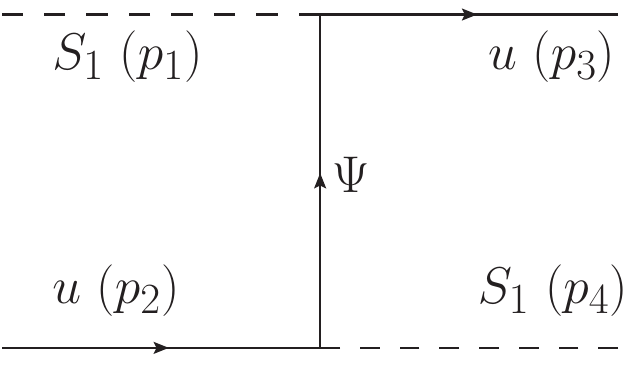}}
\subfloat[$\mathcal{M}_3$] {\label{DD_h} \includegraphics[width=0.32\textwidth]{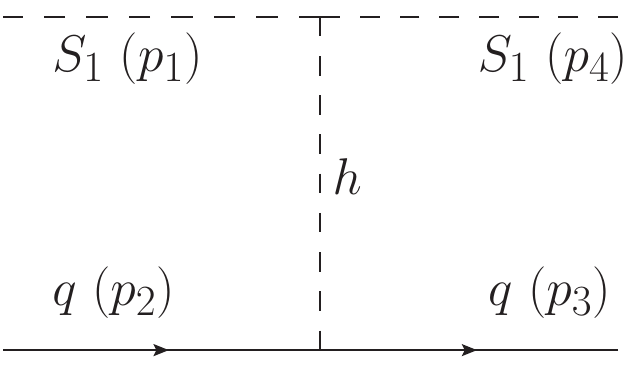}}
\caption{Scattering diagrams between scalar dark matter and the nucleon.}
\label{directDetection_5}
\end{figure}
Three different channels (Figure~\ref{directDetection_5}) are possible at the tree level for the scattering process $S_1(p_1)~u(p_2) \rightarrow S_1(p_4)~u(p_3)$, VLQ-mediated s-channel, VLQ-mediated t-channel, and Higgs-mediated t-channel diagrams. The total cross section comprises the amplitude square of the individual channels and the interference between different diagrams. The interference with different diagrams and the amplitude square of the individual diagrams are provided below. \\
Amplitude square of VLQ-mediated s-channel diagram:
 \begin{equation}
\mathcal{M}_1^\dagger \mathcal{M}_1 = \dfrac{f^4}{4}~ \dfrac{\mathcal{N} + m_u^2~(p_1.p_2+p_1.p_3)}{[(p_1+p_2)^2- M_\Psi^2~]^2}
\label{DD_ss}
\end{equation}
Amplitude square of VLQ-mediated t-channel diagram:
 \begin{equation}
\mathcal{M}_2^\dagger \mathcal{M}_2 = \dfrac{f^4}{4}~ \dfrac{\mathcal{N} - m_u^2~(p_1.p_2+p_1.p_3)}{[(p_3-p_1)^2- M_\Psi^2~]^2}
\label{DD_tt}
\end{equation}
Interference between VLQ-mediated s and t-channel diagrams:
 \begin{equation}
2\mathcal{M}_1^\dagger \mathcal{M}_2 = -2\times\dfrac{f^4}{4}~ \dfrac{\mathcal{N} + m_u^2~(-p_1.p_2+p_1.p_3)}{[(p_1+p_2)^2- M_\Psi^2~][(p_3-p_1)^2- M_\Psi^2~]}
\label{DD_st}
\end{equation}
Where $p_2^2=p_3^2=m_u^2$, $p_1^2=p_4^2=M_{S_1}^2$, and $\mathcal{N}$ is given below.
\begin{equation}
\mathcal{N} = 2(p_1.p_3) (p_1.p_2)+M_{S_1}^2~(p_1.p_3-p_1.p_2-m_u^2)+m_u^4
\label{Eq_num_5}
\end{equation}
Amplitude square of Higgs-mediated t-channel diagram:
 \begin{equation}
\mathcal{M}_3^\dagger \mathcal{M}_3 = 2 m_q^2 \lambda_{SH}^2 \cos^2\theta ~ \dfrac{p_1.p_3-p_1.p_2 - 2m_u^2}{[(p_4-p_1)^2- M_h^2~]^2}
\label{DD_hh}
\end{equation}
Interference between VLQ-mediated s-channel and Higgs-mediated t-channel diagrams:
 \begin{equation}
2\mathcal{M}_1^\dagger \mathcal{M}_3 = 2 m_u^2 \lambda_{SH} \cos\theta f^2 ~ \dfrac{p_1.p_2 + m_u^2}{[(p_1+p_2)^2- M_\Psi^2~][(p_4-p_1)^2- M_h^2~]}
\label{DD_hs}
\end{equation}
Interference between VLQ-mediated t-channel and Higgs-mediated t-channel diagrams:
 \begin{equation}
2\mathcal{M}_2^\dagger \mathcal{M}_3 = - ~2 m_u^2 \lambda_{SH} \cos\theta f^2 ~ \dfrac{p_1.p_3 - m_u^2}{[(p_3-p_1)^2- M_\Psi^2~][(p_4-p_1)^2- M_h^2~]}
\label{DD_ht}
\end{equation}

\section{Light exotic quark mass}
\label{appen_60}
The KSVZ model is characterized by the following Lagrangian, as expressed in Eq.~\ref{Lag.VLQ_5}.
\begin{equation}
\mathcal{L}\supset  f_{i}S \overline{\Psi}_L {u_i}_R+ \eta \overline{\Psi}_L \Psi_R + h.c.
\label{Lag.VLQ_50}
\end{equation}
Peccei-Quinn charge, $PQ(\Psi_L)=1$, $PQ(\Psi_R)=-1$, $PQ(\eta)=2$. In the framework of Model Eq.~\ref{Lag.VLQ_50}, the mass of the exotic quark $M_\Psi$ is expected to be approximately around the PQ breaking scale (vev of the $\eta$) or, at most, a few orders of magnitudes lower. Tuning the Yukawa coupling at least six orders of magnitude is necessary to attain an $M_\Psi$ at the TeV scale. Here, we introduce a variation where the mass of the exotic $\Psi$ quark is decoupled from the PQ scale, enabling it to be lighter. \\

Following article~\cite{Alves:2016bib}, consider another exotic quark $\Psi^{'}$. This new exotic quark $\Psi^{'}$ and the original exotic quark ($\Psi$) share identical quantum numbers, except their PQ charges: $PQ(\Psi^{'}_L)=PQ(\Psi^{'}_R)=-1$.
The new Lagrangian has the following form
\begin{equation}
\mathcal{L}\supset f_{i}S \overline{\Psi^{'}}_L {u_i}_R+ \eta \overline{\Psi}_L \Psi_R + \eta \overline{\Psi}_L \Psi_R^{'} + \overline{\Psi^{'}}_L \Psi_R  + \overline{\Psi^{'}}_L \Psi_R^{'}     + h.c.
\label{Lag.VLQ_51}
\end{equation}
After PQ breaking we get,
\begin{equation}
\mathcal{L}\supset  M_{\Psi\Psi} \overline{\Psi}_L \Psi_R + M_{\Psi\Psi^{'}} \overline{\Psi}_L \Psi_R^{'} +\tilde{M}_{\Psi^{'}\Psi} \overline{\Psi^{'}}_L \Psi_R  + \tilde{M}_{\Psi^{'}\Psi^{'}}\overline{\Psi^{'}}_L \Psi_R^{'}     + h.c.
\label{Lag.VLQ_52}
\end{equation}
Masses represented by tilde symbols can, in principle, be significantly smaller than the PQ scale. We can write 
\begin{equation}
\mathcal{M}=
\begin{pmatrix}
\tilde{M}_{\Psi^{'}\Psi^{'}} & \tilde{M}_{\Psi^{'}\Psi}\\
M_{\Psi\Psi^{'}} & M_{\Psi\Psi}
\end{pmatrix}
\end{equation}
For $\tilde{M}_{\Psi^{'}\Psi^{'}},  \tilde{M}_{\Psi^{'}\Psi} << M_{\Psi\Psi^{'}}, M_{\Psi\Psi}$, $U_L$ diagonalizing $\mathcal{M} \mathcal{M}^\dagger$ results in a small mixing angle, whereas $U_R$ diagonalizing $\mathcal{M} \mathcal{M}^\dagger$ results in a large mixing angle.\\

Upon integrating out the heaviest state, we are left with a lighter exotic quark having a mass $M_\Psi^{'}\sim \mathcal{O}(\tilde{M}_{AB})$, $A,B = \Psi, \Psi^{'}$ that exhibits a significant coupling to the Standard Model quarks through the first term of Eq.~\ref{Lag.VLQ_51}.

\bibliographystyle{JHEP}
\bibliography{ref.bib}
\end{document}